\documentclass[]{aa} \usepackage{graphics}
\def\lsim{\raise0.3ex\hbox{$<$}\kern-0.75em{\lower0.65ex\hbox{$\sim$}}}
\def\gsim{\raise0.3ex\hbox{$>$}\kern-0.75em{\lower0.65ex\hbox{$\sim$}}}
\begin{document}
\thesaurus{02(12.03.3; 08.19.4; 08.06.2; 02.14.1)}   
\title{Rates and redshift distributions of high-$z$ supernovae}  
\author{Tomas Dahl\'{e}n \and Claes Fransson}   
\institute{Stockholm Observatory,
SE-133 36 Saltsj\"{o}baden, Sweden}  
\offprints{T. Dahl\'{e}n}
\mail{tomas@astro.su.se, claes@astro.su.se} 
\date{Received / Accepted}
\titlerunning{High-$z$ Supernovae} \maketitle
\begin{abstract}
Using observed star formation rates at redshifts up to $z\sim$ 5, we
calculate cosmic supernova rates for core collapse and Type Ia
supernovae. Together with supernova statistics and detailed light
curves, we estimate the number of supernovae, and their distribution
in redshift,  that should be detectable in different filters with
various instruments, including both existing and future telescopes, in
particular the NGST.

We find that the NGST should detect several tens of core collapse
supernovae in a single frame. Most of these will be core collapse
supernovae with $1\ \lsim z\ \lsim$ 2, but about one third will have
$z\ \gsim$ 2. Rates at $z\ \gsim$ 5 are highly uncertain. For ground
based 8-10 m class telescopes we predict $\sim 0.1$ supernova per
square arcmin to I$_{AB}$ = 27, with about twice as many core collapse
SNe as Type Ia's. The typical redshift will be $z\sim$ 1, with an
extended tail up to $z \sim$ 2. Detectability of high redshift
supernovae from ground is highly sensitive to the rest frame UV flux
of the supernova, where line blanketing may decrease the rates
severely in filters below 1 $\mu$m.

In addition to the standard 'Madau' star formation rate, we discuss
alternative models with flat star formation rate at high redshifts.
Especially for supernovae at $z\ \gsim$ 2 the rates of these models
differ considerably, when seen as a function of redshift. An advantage
of using SNe to study the instantaneous star formation rate is that
the SN rest frame optical to NIR is less affected by dust extinction
than the UV-light. However, if a large fraction of the star formation
occurs in galaxies with a very large extinction the observed SN rate
will be strongly affected. An additional advantage of using SNe is
that these are not sensitive to selection effects caused by low
surface brightness. 

Different aspects of the search strategy is discussed, and it is
especially pointed out that unless the time interval between the
observations spans at least 100 days for ground based searches, and
one year for NGST, a large fraction of the Type IIP supernovae will be
lost. Because of the time delay between the formation of the
progenitor star and the explosion, observations of $z\ \gsim$ 1 Type
Ia supernovae may distinguish different progenitor scenarios.

A major problem is the determination of the redshift of these faint
supernovae, and various alternatives are discussed, including
photometric redshifts. In practice a reliable classification based on
either spectroscopy or light curves requires the SNe to be $\sim$ 2
magnitudes above the detection limit. The uncertainties in the
estimates are discussed extensively. We also discuss how the estimated rates depend on cosmology. Finally, some comments on effects
of metallicity are included. 
\end{abstract}
\section{Introduction}
The cosmic star formation rate (SFR) has now been estimated up to
redshifts $z\ \sim$ 5. By combining the evolution of the 2800 \AA\
luminosity density calculated from CFRS-galaxies (Lilly et al. 1996)
at redshifts $z\ \lsim$ 1, and the 1500 \AA\ luminosity density
calculated from high redshift galaxies ($z>$ 2) in the HDF found by
the Lyman dropout techniques, Madau et al.\ (1996) argue that the SFR
should peak at 1 $\lsim\ z\ \lsim$ 2. Complementary to these, Connolly
et al. (1997) have used HDF photometric measurements, together with
ground based near-IR photometry, to derive the 2800 \AA\ luminosity
density at redshifts 0.5 $\lsim\ z\ \lsim$ 2. These results are in
good agreement in the region overlapping with the CFRS, supporting the
case of a peak in the SFR. This conclusion is, however, sensitive to
the fact that the rest frame light may have been strongly attenuated
by dust. Absorption by dust is especially severe for the UV-light, and
since shorter rest frame wavelengths are sampled at higher $z$, this
uncertainty increases with redshift.

The evolution of the SFR is reflected in the cosmic supernova rate
(SNR). It should therefore, in principle, be possible to use supernova
(from now on SN) observations to distinguish between various star
formation scenarios. Even more important, core collapse SNe,
i.e. Types II and Ib/c, provide a direct probe of the metallicity
production with cosmic epoch. In reality these relations are
non-trivial to establish. When it comes to core collapse SNe, the
observational constraints at high redshifts are nearly
non-existent. Even at low redshift the statistics are severely affected
by selection effects. A main problem comes from the fact that core
collapse SNe observationally show a large diversity, both in terms of
luminosities and types, and with uncertain distributions. In addition,
dust absorption, as well as background contamination, affect the
statistics. Nevertheless, because of their importance for the
nucleosynthesis, as well as galaxy formation, direct observations of
the rate of core collapse SNe are  of high interest. It is therefore
hardly surprising that this is one of the main goals for the Next
Generation Space Telescope (NGST) (Stockman 1997).
 
For Type Ia SNe, the unknown time delay between formation and
explosion of the progenitors unties the link to the SFR, making
predictions more model dependent. Observations of the Type Ia rate at
high redshift therefore provides a possibility to distinguish
different progenitor scenarios.

In this paper we present estimates for the expected number of
observable SNe for the NGST, as well as for ground based instruments,
and discuss various complications entering the analysis. Previous
studies include Madau et al. (1998a), Ruiz-Lapuente \& Canal (1998),
J\o rgensen et al. (1998), Miralda-Escud\'{e} \& Rees (1997), Sadat et
al. (1998), Yungelson \& Livio (1997). With respect to most of these,
our work differs in that we include information about the light curve,
as well as spectral evolution, which allow us to predict the
simultaneously observable number of SNe. That this is important is
obvious from the fact that a nearby SN seen at the tail of the light
curve is indistinguishable from a more distant object at the peak. We divide
the SNe into different types with maximum absolute magnitudes and
spectral distributions that varies with time and type. This also
introduces a large dispersion in magnitudes at a given
redshift. Neglect of these effects introduces a severe Malmquist bias. We calculate the counts for different broad band
filters, and include information about the expected redshift
distribution of the detected SNe. Some preliminary results were given
in Dahl\'{e}n \& Fransson (1998).

Sect. 2 describes our model. Results are presented in Sect. 3. In
Sect. 4 we discuss alternative star formation scenarios. In Sect. 5 we discuss how other cosmologies affect out results. The effects of 
gravitational lensing are discussed in Sect. 6. Problems concerning 
redshift determination are discussed in Sect. 7. A general discussion 
follows in Sect. 8, and
conclusions are given in Sect. 9. Throughout most of the paper we assume a flat cosmology with
$H_0=50~{\rm km~s^{-1}~Mpc}^{-3}$ and $\Omega_M$ = 1, unless otherwise stated.
\section{The Model}
\subsection{Filters and magnitude system}
Because of the strong UV deficiency in the spectrum of most SN types,
filters bluer than R are of little interest for high-$z$ studies. For
R ($\lambda$ = 0.65 $\mu$m) and I ($\lambda$ = 0.8 $\mu$m) we use
Cousins filters, and in the near-IR J ($\lambda$ = 1.2 $\mu$m) and
K$^{\prime}$ ($\lambda$ = 2.1 $\mu$m) filters. For the M band we use a
filter that we denote by M$^{\prime}$, which is centered on $\lambda$
= 4.2 $\mu$m with $\lambda /\Delta\lambda$ = 3. For the magnitudes we
use the AB-system where m$_{AB}$ = -2.5$\log F_{\nu}$ - 48.6 (Oke \&
Gunn 1983) ($F_{\nu}$ in ergs cm$^{-2}$ Hz$^{-1}$ s$^{-1}$). Here 1
nJy corresponds to m$_{AB}$ = 31.4. When needed, we have used the
following relations to transform from Vega based magnitudes to AB
magnitudes: K$^{\prime}$ = K$^{\prime}_{AB}$ - 1.84, J = J$_{AB}$ -
0.82, I = I$_{AB}$ - 0.42 and R = R$_{AB}$ - 0.17.
\subsection{Star formation rates}
\begin{figure}
\resizebox{6.0cm}{!}{\includegraphics{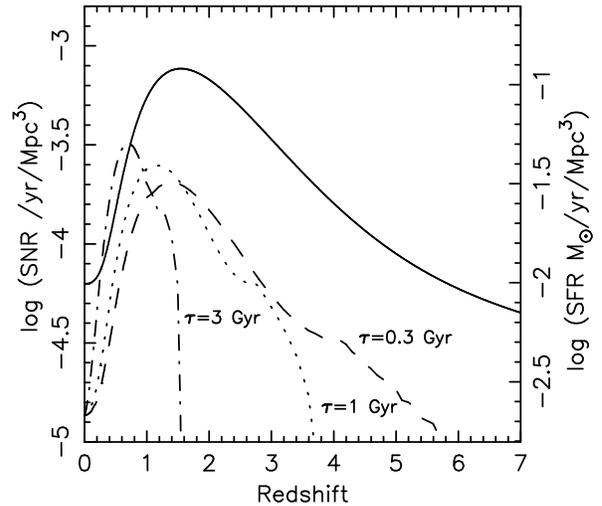}} \hfill \vskip1mm
\parbox[b]{8.5cm}{
\caption{ SNRs for the model with dust extinction $E_{B-V}$=0.1. The
solid line represent the rate of core collapse SNe. This scales
directly with the SFR, given by the right hand axis. Dashed, dotted
and dash-dotted lines represent the rate of Type Ia SNe with times
delays $\tau$=0.3, 1.0 and 3.0 Gyr, respectively. The rates of Type Ia
SNe are normalized to fit the locally observed rates.}
\label{Fig1}}
\end{figure}

A problem when using UV-luminosity densities to calculate the SFR is
that these can be matched to almost any SFR, ranging from a strongly
peaked to a flat, or even increasing, SFR at $z\geq$ 1, by adjusting
the assumed extinction due to dust. By simultaneously using luminosity
densities observed in different wavebands it is possible to break this
degeneracy. Madau (1998) used three bands (UV, optical and NIR) in
order to derive a SFR that can simultaneously reproduce the evolution
of the different luminosity densities. He found a best fit for a
universal extinction $E_{B-V}$=0.1 with SMC-type dust. Despite
corrections for absorption, the SFR still shows a pronounced peak at 1
$<z<$ 2. This SFR is shown in Fig. \ref{Fig1}. A peaked SFR is
predicted in scenarios where galaxies form hierarchically (e.g., Cole
et al. 1994). Although commonly used, this SFR is not universally
accepted and in Sect. 4 we discuss alternative SFR's.

Core collapse SNe, which originate from short lived massive stars
(ages $\lsim 5\times 10^7$ yrs), have an evolution that closely
follows the shape of the SFR. Assuming an immediate conversion of
these stars to SNe renders a multiplicative factor, $k$ = SNR/SFR, 
between the SFR
($M_{\odot} $yr$^{-1} $Mpc$^{-3}$) and the SNR (yr$^{-1} $Mpc$^{-3}$),
where $k$ depends on the IMF and the mass range of the
progenitors. Here, we take
\begin{equation}
k=\frac{\int^{50M_{\odot}}_{8M_{\odot}}\Phi (M)dM}
{\int^{125M_{\odot}}_{0.1M_{\odot}}M\Phi (M)dM}
\end{equation}
Using a Salpeter IMF yields $k$=0.0064. Note that, for consistency,
the same IMF as assumed when deriving the SFR from the luminosity
densities should to be used. Using a Scalo IMF, instead of a Salpeter
IMF, decreases the conversion  factor $k$ between the SFR and the SNR
by a factor 2.6. A Scalo IMF, however,  also increases the SFR as
derived from the UV luminosities by a factor $\sim$ 2, hence
cancelling most of the effect. The reason is that the same stars that
produce the UV luminosities also explode as core collapse SNe.

The lower mass limit for Type II progenitors is generally believed to
be 8-11 $M_{\odot}$ (e.g., Timmes et al. 1996). Here we choose 8
$M_{\odot}$, a limit supported by e.g., Nomoto (1984). An increase to
11 $M_{\odot}$ would decrease $k$, and hence the SNR, by $\sim$
38\%. Indications from especially the oxygen/iron ratio that the most
heavy stars form black holes and do not result in SNe, justifies the
use of 50 $M_{\odot}$ as an upper limit to the progenitor mass
(Tsujimoto et al. 1997). Other studies (e.g., Timmes et al. 1995),
however, find that stars with masses down to $\sim$ 30 $M_{\odot}$ may
result in black holes. An upper limit of 30 $M_{\odot}$, instead of 50
$M_{\odot}$, decreases $k$ by $\sim$ 9\%.

The uncertainties concerning the origin of Type Ia SNe makes the
relation  between the rate of these SNe and the star formation more
ambiguous. The  predicted SNR depends on the nature of the SN
progenitors. Yoshii et al. (1996) argue that the SNe Ia progenitor
lifetime is probably restricted to 0.5-3 Gyr. This range opens a
possible way to observationally distinguish between progenitor
models. Ruiz-Lapuente \& Canal (1998) find that the more short-lived
double-degenerate progenitor systems and the long-lived
cataclysmic-like systems should yield  significantly different rates.

In this paper we use as a standard model the SFRs derived by Madau
(1998) to calculate the rates of both core collapse and Type Ia
SNe. We use a universal extinction $E_{B-V}$=0.1, together with
SMC-type dust and a Salpeter IMF. In Sect. 4 we investigate how a higher
extinction affects the counts, and if it is possible to use these
counts to estimate the amount of dust.
\subsubsection{Core Collapse SNe}
The SNR used is shown in Fig. \ref{Fig1}. This is the intrinsic rate
of  exploding SNe, irrespective of magnitudes and spectral
distributions. The  apparent magnitude of a SN at redshift $z$ at a
time $t$ off its peak magnitude (i.e. $t$ can be negative) in a host
galaxy with inclination $i$ observed in a filter $f$ is given by
\begin{equation}
m_f(z,t,i)= M_f(t) + \mu(z) + K_f(z,t) + A_{g,f} + <A_{i,f}>.
\end{equation}
Here, $M_f(t)$ is the absolute magnitude of the SN in filter $f$ at
time $t$  relative to the peak of the light curve, $\mu (z)$ is the
distance modulus,
\begin{equation}
\mu(z)=5\log d_L(z)-5.
\end{equation}
The luminosity distance, $d_L$, is given by,\\

\begin{math}
d_L=\frac{(1+z)}{H_0\mid \Omega_k\mid ^{1/2}}{\rm sinn}\{ \mid
\Omega_k \mid ^{1/2}\times
\end{math} 
\begin{equation}
\times \int ^z_0
[(1+z)^2(1+\Omega_Mz)-z(2+z)\Omega_{\Lambda}]^{-1/2}dz\}.
\end{equation}
where $\Omega_k = 1 - \Omega_M -\Omega_{\Lambda}$ and 'sinn' stands
for $\sinh$ if $\Omega_k >$ 0 and for $\sin$ if $\Omega_k <$ 0 (Misner
et al. 1973). If $\Omega_k$ = 0 then the 'sinn' and the $\mid \Omega_k
\mid ^{1/2}$ terms are set equal to one. For the standard CDM
cosmology mainly used here, the distance modulus is given by $\mu (z)$= 45.4
-5log$(H_0/50$ km\ s$^{-1}$\ Mpc$^{-1}$) + 5log[(1+$z$) -
$(1+z)^{1/2}$ ]. Further, $K_f(z,t)$ gives the K-correction, $A_{g,f}$
is the Galactic absorption, and $<A_{i,f}>$ is the radially averaged
absorption in the parent galaxy with inclination $i$.

For Type Ib/c, plateau Type IIP and linear Type IIL SNe we use peak
magnitudes given by Miller \& Branch (1990). The magnitudes of the
faint SN1987A-like SNe are not well known. Here we adopt the
magnitudes given by Cappellaro et al. (1993), while magnitudes from
Patat et al. (1994) are adopted for the Type IIn SNe. The magnitudes
given by these authors are, however, not corrected for
absorption. Adopting an average $E_{B-V}$ = 0.1 yields a mean $A_B$ =
0.41 mag. Taking the effect of the albedo of the dust grains into
account lowers the effective absorption. We adopt an absorption 
$A_B$ = 0.32 mag, which is
consistent with the mean face-on absorption calculated by Hatano et
al. (1998) in their models for Type II SN extinction. We have here
assumed that the absorption of the light from the parent galaxy and
the SNe follow the same extinction law, implying that the SNe and the
progenitor stars occur in the same environment, which to some extent
may be incorrect. If the first core collapse SNe and their main
sequence progenitors in a region sweeps away part of the shrouding
dust, the absorption could be lower than calculated above. In Sect. 4 we
return to this possibility. In Table \ref{table1} we list the
corrected mean absolute B magnitudes, as well as the adopted
dispersion in this quantity for the different types at the peak of the
light curve.
\begin{table}
\begin{center}
\begin{tabular}{lccc}
\hline \bf SN Type & \bf $<M_{B,0}>$ & \bf $\sigma$ & $f$ \\ \hline
SNIb/c & -18.44 & 0.39 & 0.23\\ SNIIP & -17.86 & 1.39 & 0.30\\ SNIIL &
-18.38 & 0.51 & 0.30\\ SN87A-like & -15.43 & 1.0 & 0.15\\ SNIIn &
-20.03 & 0.6 & 0.02\\ \hline
\end{tabular}
\caption{ Adopted maximum absolute B magnitudes and dispersion in the
AB-system.  From Miller \& Branch (1990), Cappellaro (1997),
Cappellaro et al. (1993) and  Patat et al. (1994). Values are
corrected for extinction $E_{B-V}$=0.1,  corresponding to
$A_B$=0.32. The intrinsic fraction of exploding SNe of the  different
types are given by $f$.}
\label{table1}
\end{center}
\end{table}

For the evolution of the luminosity with time we use light curves from
Filippenko (1997) for the Type Ib/c's, IIP's, IIL's and the
SN1987-like SNe. We have made some modifications to the light curve
for the IIP's to achieve a better fit to the observed, as well as
theoretical, light curves presented in Eastman et al. (1994). For Type
IIn's we use a light curve intermediate between IIP and IIL. This
seems adequate in view of the data presented by Patat et al. (1994),
who find that Type IIn SNe can have both linear and plateau shape, but
there are also light curves in-between these two.

The K-correction is calculated by assuming modified blackbodies for
the spectral distributions of the SNe. At $z\gsim$ 1 even the I-band
corresponds to the rest UV part of the spectrum. The exact spectral
distribution in the blue and UV is therefore crucial. Unfortunately,
this part of the spectrum is relatively unexplored even at low $z$. UV
observations of Type IIP's are especially scarce. Only for the
somewhat peculiar Type IIP SN1987A is there a good UV coverage (Pun et
al. 1995). This showed already $\sim$ 3 days after explosion a very
strong UV deficit, similar to Type Ia SNe. This is a result of the
strong line blanketing by lines from Fe II, Fe III, Ti II and other
iron peak elements (e.g., Lucy 1987; Eastman \& Kirshner
1989). Although the progenitors of typical Type IIP's probably are red
supergiants, rather than blue as for SN1987A, there is no reason why
these ions should be less abundant than in SN1987A. On the contrary,
because of the near absence of strong circumstellar interaction and
similar temperature evolution they are expected to have fairly similar
UV spectra, as calculations by Eastman et al. (1994) also
show. Eastman et al. find that the UV blanketing sets in after $\sim$
20 days when the effective temperature becomes less than $\sim$ 7000
K. This coincides with the beginning of the plateau phase. We
therefore mimic the UV blanketing by a 2 magnitude drop between 4000
\AA\ and 3000 \AA , and a total cutoff short-wards of 3000 \AA , for
the Type IIP and SN1987A-like SNe with T$_{eff}\lsim$ 7000 K. At
higher T$_{eff}$ we assume non-truncated blackbodies for these
types. The time evolution of the spectra is modeled by changing the
characteristic temperature of the blackbody curves to agree with the
calculations by Eastman et al. At shock outbreak a short interval of
energetic UV radiation occurs with $10^5$ $\lsim$ T$_{eff}$ $\lsim$
$10^6$ K. This only lasts a few hours, but may due to its high
luminosity be observable to high $z$ (Klein et al. 1979; Blinnikov 
et al. 1998; Chugai et al. 1999). It then cools to $\sim$ 2.5$\times 
10^4$ K at $\sim$ 1 day and
further to $\lsim$ 7000 K after $\sim$ 20 days. At the plateau phase
the temperature is $\sim$ 5000 K.

From the limited UV information of the Type IIL's SN 1979C, SN 1980K
and SN 1985L (Cappellaro et al. 1995), and the Type IIn SN 1998S
(Kirshner et al., private communication), we model the spectra of the
IIL's and IIn's by blackbody spectra without cutoffs. The most
extensive coverage of the UV spectrum of a Type IIL exists for SN
1979C (Panagia 1982). By fitting blackbody spectra to the optical
photometry, we find that even after two months, when the color
temperature is only $\sim$ 6000 K, there is no indication of UV
blanketing. On the contrary, there is throughout the evolution a
fairly strong UV excess, consisting of continuum emission as well as
lines, in particular Mg II 2800 \AA. The UV excess of these SNe is
likely to be caused by the interaction of the SN and its circumstellar
medium. This ionizes and heats the outer layers of the SN, decreasing
the UV blanketing strongly. For the effective temperature as a
function of time we use the fits for SN 1979C by Branch et al. (1981). 
The Type Ib/c light curve is taken from Filippenko (1997) and is
assumed to have the same  spectral characteristics as the Type Ia
(described further below).

The Galactic absorption, $A_{g,f}$, is taken to be zero in our
modeling. This can easily be changed to other values of $A_{g,f}$. The
term $<A_{i,f}(z)>$ adds the absorption due to internal dust in the
parent galaxy with inclination $i$, according to the adopted
extinction laws. Gordon et al. (1997) show that the extinction curve
for starburst galaxies lacks the 2175 \AA\ bump, like an SMC-type
extinction curve does, and shows a steep far-UV rise, intermediate
between a Milky Way and an SMC-like extinction curve. The observed
increase with redshift of the UV-luminosity originates mainly from
starburst/irregular systems (e.g., Brinchmann et al. 1998). This
implies that a major part of the core collapse SNe should be found in
such galaxies, and an SMC-type extinction curve should therefore be
most appropriate when calculating the absorption of the SNe light. The
difference between an SMC-type dust curve and a Milky Way-type
extinction in the interesting wavelength range is small. This is
especially true for SNe with a short wavelength cutoff in their
spectral energy distributions, but also a blackbody spectrum with
T$_{eff}\lsim$ 7000 K drops fast enough at short wavelengths for the
precise form of UV absorption in this region to be less important.

The dependence on inclination has been modeled by Hatano et
al. (1998). The absorption closely follows a (cos $i$)$^{-1}$ behavior
up to high inclinations. We approximate their results by $<A_{B,i}>$ =
0.32[(cos$i$)$^{-1}$-1]. Also the radial dependence of the absorption
is discussed by Hatano et al. We simplify our calculations by adopting
their radially averaged value of the absorption. Dividing the host
galaxies into different types increases the dispersion in
absorption. Finally, the extinction may depend on the intrinsic
luminosity and the metallicity of the host galaxy. This variation
should to some extent already be accounted for in the observed
dispersion of the peak magnitudes.

The SNe are divided into the five different groups, with fractions,
$f$, representing the intrinsic fraction of exploding core collapse
SNe of the different types (i.e. irrespective of magnitude). We
estimate $f$ by using the observed {\it ratios of discovery},
$f_{obs}$, given by Cappellaro et al. (1993), who find
$f_{obs}$(IIP)$\simeq f_{obs}$(IIL), $f_{obs}$(Ib/c)$\simeq$
0.3$f_{obs}$(II) and $f_{obs}$(IIn)$\simeq$ 0.2$f_{obs}$(II). 

The total number of Type II's is the sum of Type IIP, IIL, 1987A-like
and IIn. Adding the Type Ib/c yields the total number of core collapse
SNe. Cappellaro et al. (1997) argue that the intrinsic fractions of
IIn and 1987A-like should be $f$(IIn) = (0.02 - 0.05)$f$(II) and
$f$(1987A-like) = (0.10 - 0.30)$f$(II). The IIP, IIL and Ib Types have
approximately the same magnitudes, i.e. the ratio of discovery should
be close to the intrinsic fractions between these. Combining these
values and assumptions leads to our adopted intrinsic fractions in
Table \ref{table1}. We note here that, unfortunately, the rates of the
different classes are affected by fairly large uncertainties.

The observable number of SNe with different apparent magnitudes is
calculated by integrating the SNR over redshift. The SNe are
distributed between the different types according to their intrinsic
fractions, and are placed in parent galaxies with inclinations between
0\degr\ to 90\degr. An important feature of our model is that the
number of SNe exploding each year are distributed in time, and are
given absolute magnitudes consistent with their light curves. With
this procedure we obtain the simultaneously observable number of SNe,
including both those close to peak and those at late epoch. In order
to actually detect the SNe, at least one more observation has to be
made after an appropriate time has passed. In Sect. 8 we discuss the
spacing in time between observations.

For $z\ >$ 5 we use a rate that is an extrapolation from lower
$z$. This certainly simplifies the actual situation
drastically. However, due to the large distance modulus and the
decline in the rate at high $z$, the fraction of SNe with $z\ >$ 5 is
small, and hence the errors due to the uncertainty in the shape of the
SFR at $z\ >$ 5. Furthermore, SNe with a spectral cutoff at short
wavelengths drop out at redshifts $z \sim \lambda_{eff}/4000$ -1,
where $\lambda_{eff}$ is the effective wavelength of the filter. In
the R, I, J, K$^{\prime}$ and M$^{\prime}$ filters this occurs at $z
\sim$ 0.65, 1.0, 2.0, 4.5 and 10, respectively. This makes the
contribution from SNe with higher redshifts insignificant. A caveat
here is that lensing, as well as an early epoch of Pop III SNe, may
cause a significant deviation from this extrapolation. These issues
are discussed in Sect. 4 and Sect. 6.
\subsubsection{Type Ia SNe}
When calculating the number of Type Ia SNe we employ the same
procedure as above. We use an extinction corrected peak magnitude
$M_B$ = -19.99 and dispersion $\sigma$ = 0.27, found by Miller \&
Branch (1990). The time delay between the formation of the progenitor
star and explosion of the SNe is treated in a way similar to Madau et
al. (1998a). Most likely, the progenitors are stars with mass
$3M_{\odot}<M<8M_{\odot}$ (Nomoto et al. 1994). Stars forming at time
$t^{\prime}$ reach the white dwarf phase at $t^{\prime}+\Delta
t_{MS}$, where $\Delta t_{MS}$=10($M/M_{\odot})^{-2.5}$ Gyr is the
time spent on the main sequence. After spending a time $\tau$ in the
white dwarf phase, a fraction $\eta$ of the progenitors explode as a
result of binary accretion at $t=t^{\prime}+\Delta t_{MS}+\tau$. The
SNR at time $t$ can then be written\\

\begin{math}
SNR(t)=\eta \int^t_{t_F}SFR(t^{\prime})dt^{\prime}\times
\end{math}
\begin{equation}
\times\int^{8M_{\odot}}_{3M_{\odot}}\delta(t-t^{\prime}-\Delta
t_{MS}-\tau )\phi (M)dM,
\end{equation}
where $t_F$ is the time corresponding to the redshift of the formation
of the first stars, $z_F$. Arguments from Yoshii et al. (1996) and
Ruiz-Lapuente \& Canal (1998) indicate 0.3 $\lsim\ \tau\ \lsim$ 3
Gyr. In the calculations we therefore use three different values for
the time delay, $\tau$ = 0.3, 1 and 3 Gyr. The $\tau$ = 0.3 Gyr model
approximately mimics the double degenerate case, while the $\tau$ = 1
Gyr model resembles the cataclysmic progenitor model. With the
additional $\tau$ = 3 Gyr the range is expanded to cover all likely
models.

The parameter $\eta$ in Eq. (6) is introduced to give the fraction of
stars in the interval $3-8$ $M_{\odot}$ that result in Type Ia SNe,
and is determined by fitting the estimated SNR at $z$ = 0 to the
locally observed Type Ia rates. For an alternative approach based on
specific assumptions about the progenitors see J\o rgensen et
al. (1997). Results from local SN searches (Cappellaro et al. 1997;
Tammann et al. 1994; Evans et al. 1989) give local rates of 0.12,
0.19, and 0.12 SNu, respectively (1 SNu= 1 SN per century per
10$^{10}L_{B\odot}$). Adopting a mean of 0.14 $\pm$ 0.06 SNu (Madau et
al. 1998a), and using a local B band luminosity estimated by Ellis et
al. (1996), leads to a local Type Ia rate of 1.3 $\pm$ 0.6
$\times$10$^{-5}$ SNe yr$^{-1}$ Mpc$^{-3}$. The normalization at $z$ =
0 yields an efficiency 0.04 $<\eta<$ 0.08, where the range is due to
the fact that different values of $\tau$ gives different $\eta$ in
order to reproduce the local rates. The uncertainty in the local SNR
is equivalent to an additional spread in $\eta$ by a factor $\sim$
3. The normalization to the local rate leads to an uncertainty in the
Type Ia SNR at all redshifts that corresponds to the uncertainty in
the local rate. This implies an uncertainty by a factor $\sim$ 3 in
the estimated Type Ia rates.

The treatment of absorption in the case of Type Ia SNe is more
complicated than in the case of core collapse SNe, and is therefore
subject to larger uncertainties. The long time delay between the
formation of the progenitors and the explosion unties the
environmental link between these events. A (binary) star forming in a
dusty starburst region may e.g. explode much later as a Type Ia SN in
a dust-free elliptical galaxy. Observations show that Type Ia SNe do
occur in both elliptical and spiral galaxies. However, the fact that
the major part of local Type Ia SNe are detected in spirals, together
with observations that indicate that the global fraction of
ellipticals, or at least low-dust ellipticals, seems to decrease at
increasing redshifts (Driver et al. 1998) may justify our
simplification of putting all the SNe Ia in spiral
environments. Ignoring the elliptical parent galaxies leads to a
slight overestimate of the absorption, and a slight underestimate of
the SN detection rate.

Absorption in both the bulge and disk components of spiral galaxies
have been  calculated by Hatano et al. (1998). They find a somewhat
smaller absorption  for Type Ia SNe than for core collapse SNe. The
Type Ia absorption is also less dependent on the inclination of the
parent galaxy. We use the disk component results of Hatano et al. in
our model for the absorption of the Type Ia SNe. Fig. \ref{Fig1} show
the intrinsic Type Ia SNRs for the three time delays used. Increasing
the time delay shifts the peak of the Ia's towards lower redshift. The
SFR (and the rate of core collapse SNe) peak at $z\sim$ 1.55, while
the Type Ia rates in the $\tau$ = 0.3, 1 and 3 Gyr models peak at
$z\sim$ 1.35, $z\sim$ 1.16, and $z\sim$ 0.71, respectively.

To describe the spectral energy distribution of the Type Ia SNe we use
blackbody curves with a spectral cutoff at $\sim$ 4000 \AA~ (e.g.,
Branch et al. 1983). The temperature is set to 15\,000 K around the
peak, decreasing to 6000 K after $\sim$ 25 days (e.g., Schurmann
1983). We have compared the K-corrections in our model to those
calculated from real spectra by Kim et al. (1996). The mean deviation
in the R-band correction between our modified black body curves and
the detailed calculations by Kim et al. is $\Delta m \sim$ 0.1
mag. This agreement justifies the use of the blackbody representation
for the SN spectra. The average Type Ia light curve is taken from
Riess et al. (1999).
\section{Results}
Using the model above we can now estimate the observed
number of SNe per square arcmin down to some limiting magnitude,
including corrections from extinction and the shift of the spectrum
with redshift.

\begin{figure*}
\resizebox{\vsize}{!}{\includegraphics{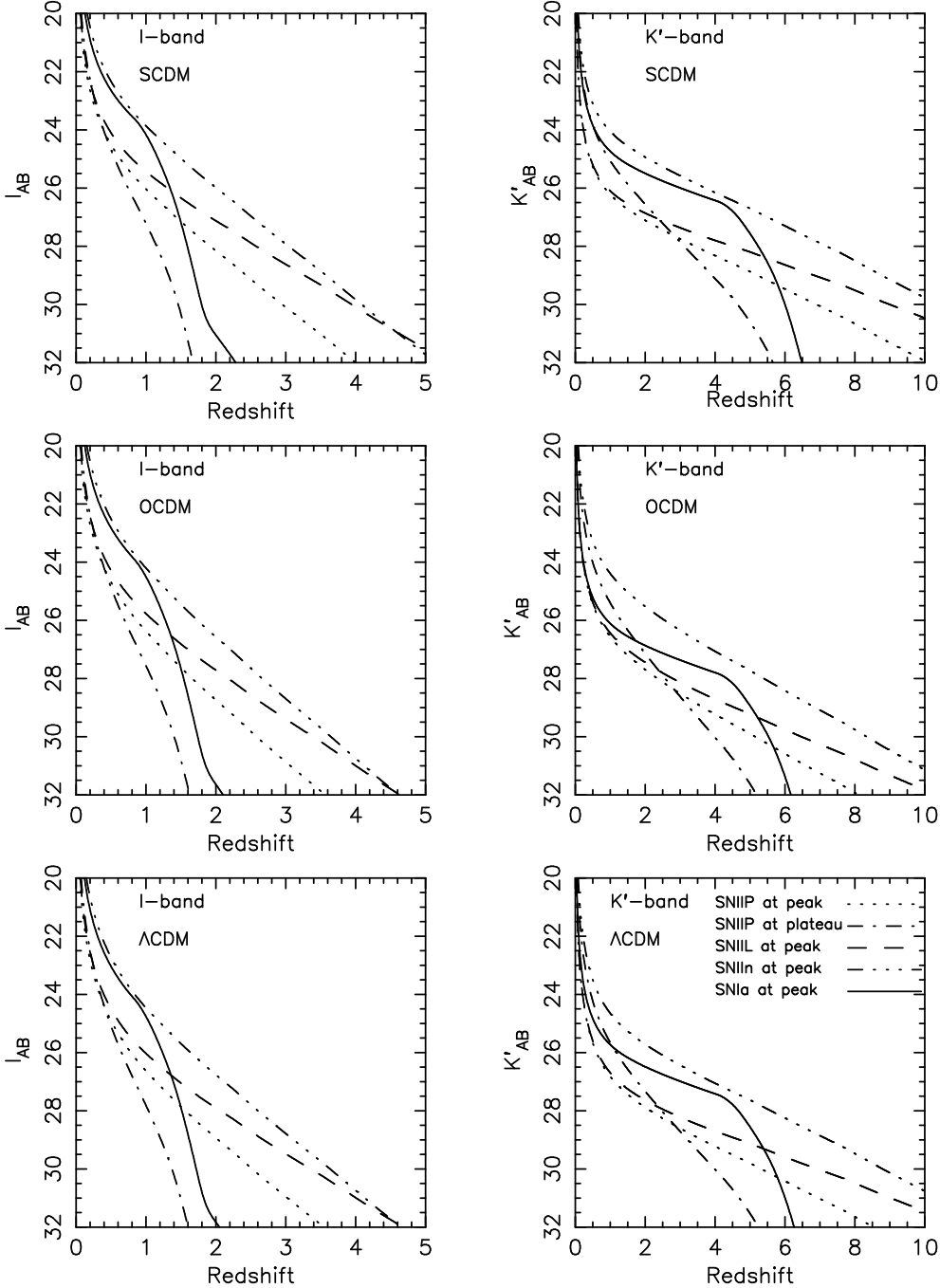}} 
\hfill 
\vskip4mm
\parbox[b]{17cm}{
\caption{ Magnitudes as a function of redshift for different SN types
in the I-band (left panels) and the K$^{\prime}$-band (right
panels). Top panels show SCDM ($\Omega_M$ = 1, $\Omega_{\Lambda}$ = 0), middle panels show OCDM ($\Omega_M$ = 0.3, $\Omega_{\Lambda}$ = 0) and lower panels show $\Lambda$CDM ($\Omega_M$ = 0.3, $\Omega_{\Lambda}$ = 0.7). Dotted line shows Type IIP at peak while the dash-dotted line
shows the Type IIP at the plateau $\sim$ 40 days after the
explosion. Dashed, dash-triple-dot and solid lines show Type IIL, Type
IIn and Type Ia, respectively. The magnitude for Type Ib/c's at peak
follows the Type Ia's, except for an offset of $\Delta$m $\simeq$ 1.5
mag towards fainter magnitudes. The magnitude of 1987A-like SNe
resembles the Type IIP at the plateau, but with an offset $\Delta$m
$\simeq$ 2.0 towards fainter magnitudes.}
\label{Fig2}}
\end{figure*}

As an illustration, we show in Fig. \ref{Fig2} the peak magnitudes (i.e. at $\sim$ 10-15 days after explosion) for Types IIP, IIL, IIn
and Ia as a function of redshift. We also show the magnitude of a Type
IIP at the plateau phase (age $\sim$ 40 days). The magnitude for Type
Ib/c's at peak follows the Type Ia's, except for an off-set of
$\Delta$m $\simeq$ 1.5 mag towards fainter magnitudes. The magnitude
of SN1987A-like SNe at the peak resembles the Type IIP at the plateau,
but with an off-set $\Delta$m $\simeq$ 2.0 towards fainter
magnitudes. Besides the standard flat $\Omega_M$ = 1 cosmology (hereafter SCDM), we also show the apparent magnitudes for an open cosmology (OCDM) with $\Omega_M$ = 0.3 and $\Omega_{\Lambda}$ = 0, and a flat, $\Lambda$-dominated cosmology ($\Lambda$CDM) with $\Omega_M$ = 0.3 and $\Omega_{\Lambda}$ = 0.7.
Note that the curves have a dispersion in magnitude
according to values given in Table \ref{table1}. The dispersion is
largest for the Type IIP's; the peak magnitude of these may vary by
more than one magnitude. Fig. \ref{Fig2} shows that the Type Ia's and
the Type IIP's at the plateau drop out of the I and K$^{\prime}$-band
at $z \sim$ 1 and  $z \sim$ 4.5, respectively, as a result of their
UV-cutoffs. In contrast, the UV-bright Type IIL's and IIn's stay
relatively bright in the I band even at high $z$.

\subsection{Core collapse rates}
\begin{figure*}
\resizebox{12.5cm}{!}{\includegraphics{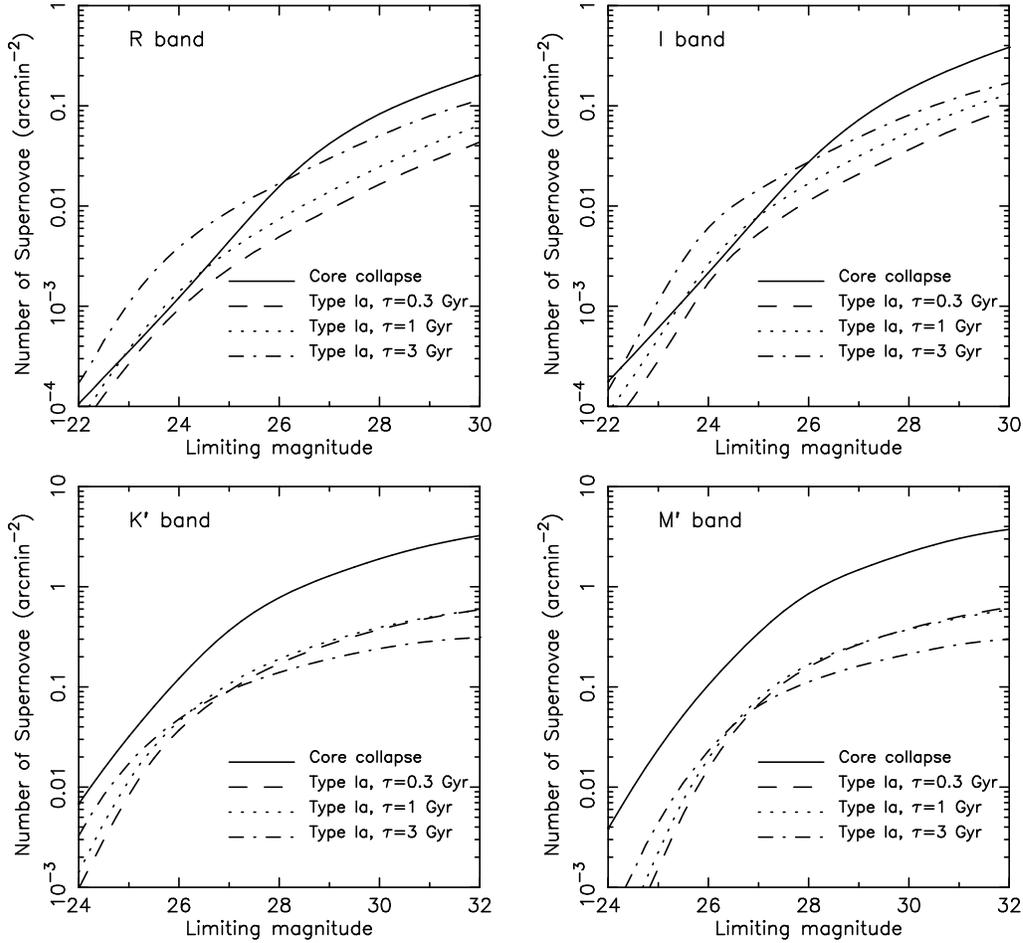}} \hfill \vskip1mm
\parbox[b]{17cm}{
\caption{Number of SNe per square arcmin that can be detected down to
different limiting magnitudes in the R, I, K$^{\prime}$ and
M$^{\prime}$ bands. Solid lines represent core collapse SNe. Dashed,
dotted and dash-dotted lines represent Type Ia SNe with time delays
$\tau$=0.3, 1 and 3 Gyr,  respectively. Note that AB-magnitudes are
used.}
\label{Fig3}}
\end{figure*}

The solid lines in Fig. \ref{Fig3} show the number of predicted core
collapse SNe per square arcmin in the R, I, K$^{\prime}$ and
M$^{\prime}$ filters for different limiting AB-magnitudes. Because of
the drop in the UV flux, bands bluer than R are of less interest for
high redshifts.

According to the specifications, NGST should have a detection limit of
$\sim$ 1 nJy in the J and K$^{\prime}$ bands and $\sim$ 3 nJy in the
M$^{\prime}$ band for a 10$^4$ s exposure with S/N=10 and $\lambda
/\Delta \lambda$ = 3 (Stockman 1997). These fluxes correspond to AB
magnitudes of J$_{AB}$ = 31.4, K$^{\prime}_{AB}$ = 31.4 and
M$^{\prime}_{AB}$ = 30.2, respectively. Using these limits, we find
that the K$^{\prime}$ band yields the highest number of core collapse
SNe. In a 4$\times$4 square arcmin field we predict $\sim$ 45
simultaneously detectable core collapse SNe, with a mean redshift
$<z>$ = 1.9.

\begin{figure*}
\resizebox{3.5cm}{!}{\includegraphics{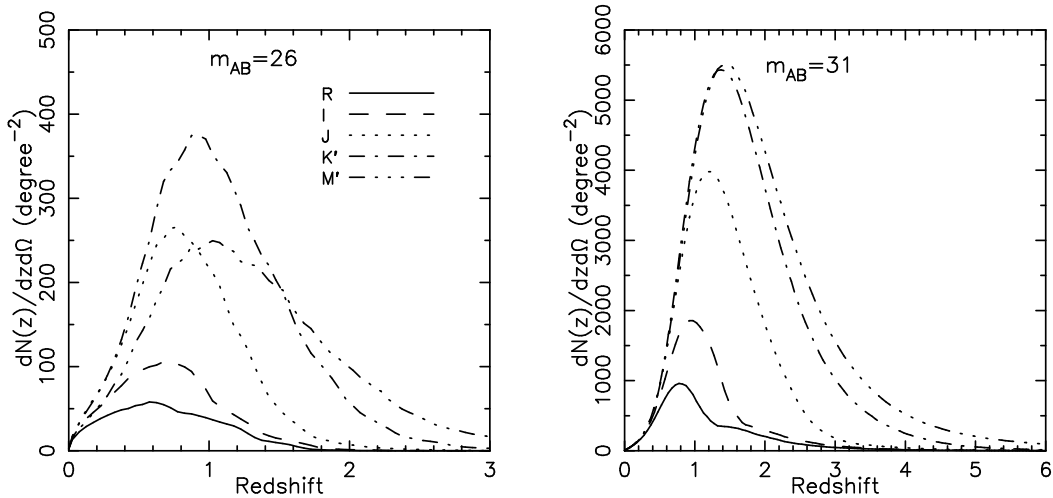}} \hfill \vskip1mm
\parbox[b]{17cm}{
\caption{Redshift distribution of core collapse SNe in the R, I, J,
K$^{\prime}$ and M$^{\prime}$ filters for  limiting magnitudes
m$_{AB}$=26 (left panel) and m$_{AB}$=31 (right panel).}
\label{Fig4}}
\end{figure*}

In Fig. \ref{Fig4} we show the redshift distribution in all bands for
m$_{AB}$ = 26 and m$_{AB}$ = 31. For the lower magnitude limit one clearly
sees the advantage of the infrared J, K$^{\prime}$ and M$^{\prime}$
bands when it comes to detect SNe with $z \gsim$ 1. This is, of course,
even more pronounced for m$_{AB}$ = 31, with the M$^{\prime}$ band
having the highest number of SNe at $z \gsim$ 2. Note, however, that
reaching R$_{AB}\sim$ 26 is considerably easier than J$_{AB}$ or
K$^{\prime}_{AB}\sim$ 26 from ground. The smaller detector size in the
latter bands is also a severe limiting factor. 

For SNe with redshifts $z>$ 5 we estimate $\sim$ 1 SN per NGST field
in the M$^{\prime}$ filter. The actual numbers of the high-$z$ SNe are
highly uncertain, since the SNR at $z>5$ is based on an extrapolation
from lower redshifts. Also gravitational lensing may be important at
these redshifts (see Sect. 6). However, independent of the actual
numbers, we find that the M$^{\prime}$ filter yields a factor $\gsim$
2-5 higher counts compared to the K$^{\prime}$ filter at these
redshifts. We also find that NGST should be well suited to detect SNe
originating from a possible Pop III at redshifts $\gsim$ 10.

The R and I bands sample light with rest wavelengths short-ward of the
peak in the blackbody curves at lower redshifts than the J,
K$^{\prime}$ and M$^{\prime}$ bands do. Besides the drop in luminosity
due to the spectral shape (an effect especially pronounced for SNe
with strong UV blanketing), these wavelengths are affected by larger
extinction. The large K-correction decreases the rates in the R and I
bands, and few SNe with $z \gsim$ 1 are detected in I and R, even for
limiting magnitudes $\gsim$ 29. Increases in these bands are instead
caused by sampling SNe with fainter absolute magnitudes at $z \lsim$ 1.

\begin{figure*}
\resizebox{12.5cm}{!}{\includegraphics{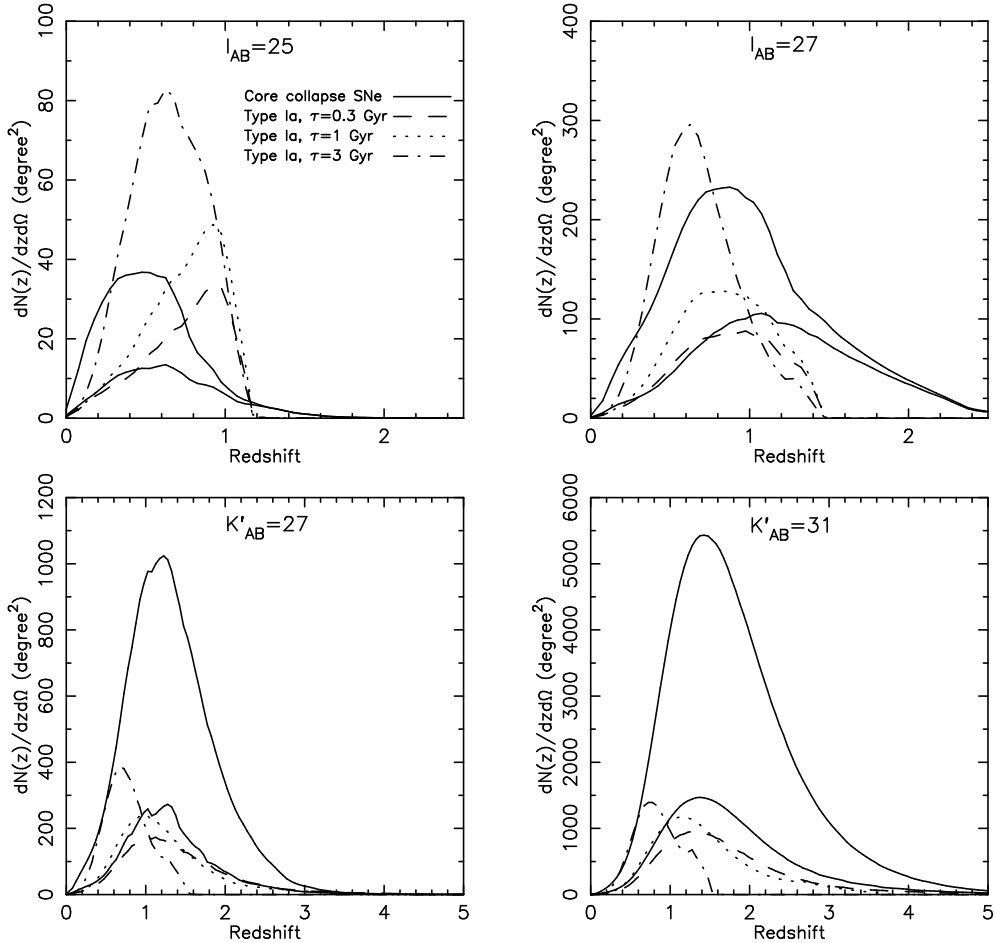}} \hfill \vskip1mm
\parbox[b]{17cm}{
\caption{Top panels: Observed SN distribution in redshift in the
I filter, for limiting magnitudes I$_{AB}$=25 and I$_{AB}$=27. Upper
solid lines show the total number of core collapse SNe, while the
lower solid lines show the part originating from Type IIL and Type IIn
SNe, i.e the SNe without a UV cutoff in their spectra over the whole
light curve. Dashed, dotted and dash-dotted lines show the number of
Type Ia SNe for models with $\tau$=0.3, 1 and 3 Gyr,
respectively. Bottom panels: Same for K$^{\prime}$ filter with
limiting magnitudes K$^{\prime}_{AB}$=27 and
K$^{\prime}_{AB}$=31. Note the expanded redshift scale in the
K$^{\prime}$ band, as well as the different scales of the y-axis.}
\label{Fig5}}
\end{figure*}

In Fig. \ref{Fig5} we show the redshift distribution of core collapse
and Type Ia SNe down to I$_{AB}$ = 27 and K$^{\prime}_{AB}$ = 31,
respectively. The total number of core collapse SNe is shown as the
solid lines in each panel. The lower solid line shows the fraction of
these coming from Type IIL and IIn SNe, which lack UV-cutoff in their
spectrum over the whole light curve. The I$_{AB}$ = 27 panel clearly
demonstrates the dominance of these at high redshift. Besides these,
only the fraction of Type IIP's which are seen near the peak, before
UV blanketing sets in, contribute to the SNe with $z \gsim$ 1.5. The
K$^{\prime}$ band is far less sensitive to this effect, since the UV
cutoff does not affect this filter at $z \lsim$ 4.5. When we compare
the redshift distribution for K$^{\prime}$ = 27 and K$^{\prime}$ = 31,
we note that the mean redshift does not change much. However, the
absolute number increases by a factor $\sim$ 5. This is mainly caused
by SNe below the light curve peak. Also the number of high redshift
SNe with $z \gsim$ 2 increases by a large factor.

Table \ref{Table2} gives some examples of our estimates for the number
of SNe for NGST, VLT/FORS and HST/WFPC2, all for an exposure of 10$^4$
s and S/N = 10. For VLT/ISAAC and HST/NICMOS the small field and the
relatively bright limiting magnitudes, compared to NGST, do not result
in more than $\sim$ 0.01 SNe per field. These instruments are
therefore of limited interest when it comes to detecting high-$z$ SNe.

So far we have discussed the number of SNe that are simultaneously
observable during one search. To actually detect the SNe, additional
observations are obviously required. Preferentially, a series of observations of
each field should be undertaken in order to obtain a good sampling of
the light curves. This also leads to the detection of new SNe in the
additional frames. The number of new SNe depends on the total length
of the search, and the spacing in time between each observation. As an
illustration, in an idealized situation where a field is covered
continuously during one year, the detected number of SNe per square
degree, with limit I$_{AB}$ = 27, is increased from $\sim$ 260 in the
first image, to $\sim$ 1650 for the whole years coverage. This
procedure is, of course, observationally unrealistic. Using 80 days
between each observation (i.e. $\sim$ five observations for a years
coverage) results in a total of 1150 different SNe. Using 40 days
instead ($\sim$ ten observations) gives $\sim$ 1400 different SNe. Due
to the time dilatation factor $(1+z)$, one observes relatively fewer
new SN explosions at high redshifts compared to the first detection.

With NGST, using limits as above, we estimate $\sim$ 45 SNe in each
frame. The total number of different SNe in a field that is covered
continuously during one year is $\sim$ 68. This means that in addition
to the $\sim$ 45 SNe observed in the first field, only $\sim$ 23 new
SNe have exploded during the year. With three observations and 180
days between the observations, $\sim$ 63 different SNe are detected.

Compared to earlier estimates, our use of complete light curves during
the whole evolution, as well as distribution of the SNe over time,
results in a larger number of SNe, as well as a realistic distribution
over redshift for a given magnitude. For example, with the same SFR
and extinction, Madau et al. (1998a) predict $\sim$ 7 SNe per NGST
field per year in the range 2 $<z<$ 4. Our calculations result in
$\sim$ 22. The difference is due to the fact that we use  light curves
covering the whole evolution, which allow us to include SNe at all
epochs, instead of only those at peak, which is the case in Madau et
al. 
\begin{table*}
\begin{center}
\begin{tabular}{|l|ccc|ccc|ccc|}
\hline Instrument& Field & Filter & limit & \multicolumn{3}{|c|}{Core
collapse}& \multicolumn{3}{|c|}{Type Ia ($\tau$=1 Gyr)}\\ &
(arcmin$^2$) &
&(m$_{AB}$)&N(tot)&N($z>1$)&N($z>$2)&N(tot)&N($z>1$)&N($z>$2)\\ \hline
NGST & 16 & M$^{\prime}$ & 30.2 & 37 & 33 & 15 & 6.3 & 4.8 & 1.2 \\
NGST & 16 & K$^{\prime}$ & 31.4 & 45 & 40 & 16 & 8.5 & 6.5 & 1.6 \\
NGST & 16 & J & 31.4 & 27 & 22 & 5& 6.3 & 4.4 & 0.6\\ VLT/FORS & 46 &
I & 26.6 & 2.4 & 0.9 & 0.05 & 1.1 & 0.3 & -\\ VLT/FORS & 46 & R & 26.8
& 1.6 & 0.7 & 0.03 & 0.57 & 0.03 & -\\ HST/WFPC2 & 5 & I & 27.0 & 0.36
& 0.16 & 0.02 & 0.15 & 0.05 & - \\ HST/WFPC2 & 5 & R & 27.5 & 0.30 &
0.15 & 0.02 & 0.09 & 0.01 & -\\ \hline
\end{tabular}
\end{center}
\caption{Estimated number of SNe per field for different
instruments. The limiting magnitudes are given for a 10$^4$s exposure
and S/N=10.}
\label{Table2}
\end{table*}
\subsubsection{Shock breakout supernovae}
Chugai et al. (1999) have noticed that the short peak in the light
curve connected with the shock breakout may give rise to a transient
event with a duration of a few hours. The possibility to observe this
was pointed out already by Klein et. al. (1979), although they
concentrated mainly on the soft X-ray range. 

Based on a radiation-hydrodynamics code, similar to that of Eastman et
al. (1994), Chugai et al. have calculated monochromatic light curves
for a Type IIP SN (or rather a scaled SN 1987A model) and a Type IIb
(specifically SN1993J). With a short time interval between
observations these SNe will be easily distinguishable from the Type Ia
and Ib/c SNe, which have a rise time of t $\gsim$ 20 days (at $z 
\gsim$ 1), and therefore only show a modest change in luminosity. A
major problem in comparing their results to ours is that it is not
discussed how they obtain their adopted intrinsic SNR's, although they
approximately agree with those used in this paper, as well as Madau
(1998). They also neglect dust extinction in their calculations.

Using this model, Chugai et al. find that two deep exposures, separated
by $\sim$ 10 days, result in 1.3 Type II SNe in the 6.8$\times$6.8
square arcmin field of the VLT/FORS camera, using limit
I$_{AB}$ = 28.2. In the calculations Chugai et al. assume that all SNe
with $z \leq$ 2 are detected with this limit.

Using our hierarchical model, which gives approximately the same SNR
up to $z \sim$ 2, and the same observational set-up, as Chugai et al.,
our calculations result in $\sim$ 0.27 SNe. The reason for our lower
estimate is that Chugai et al. assume all Type II SNe to have the same
steep initial rise as the Type IIP and IIb. In our model we do not
include any shock breakout for Types IIL and IIn, since the early
time behavior of these SN types is not well known. {\it If} we do include
all Type II SNe, our estimate increases to 0.66 SNe. The remaining
discrepancy is mainly caused by the simplification Chugai et al. do by
assuming that all SNe with $z \leq$ 2 are detected, combined with the
fact that they do not include dust extinction in their
calculations. Nevertheless, this may be an interesting way
of studying the shock breakout of SNe. Unfortunately, it may be
difficult to estimate the temperature and luminosity separately from
this type of observations, because most of the observed evolution will
be in the Rayleigh-Jeans part of the spectrum. Soft X-rays, as
proposed by Klein et al., is here a better probe.
\subsection{Type Ia SNe}
Fig. \ref{Fig3} shows that in the R and I bands the number of core
collapse SNe is comparable or larger than the number of Type Ia SNe
for magnitudes fainter than $25-26$. The exact crossing point depends
on the life time of the SN Ia progenitors, as well as the Type Ia
normalization at low $z$. In the K$^{\prime}$ and M$^{\prime}$ bands
core collapse SNe tend to dominate at all magnitudes. The reason for
this difference is that the Type Ia SNe have a higher effective
temperature over a longer period than the Type II SNe. The optical to
IR flux ratio is therefore higher for the Type Ia's. 
 
To illustrate the dependence on the progenitor life time,
Fig. \ref{Fig3} gives the number of Type Ia SNe for $\tau$ = 0.3, 1
and 3 Gyr in the different filters. Fig. \ref{Fig3} shows that a
change in the progenitor life time, $\tau$, introduces a
non-negligible variation in the predicted number of observable
SNe. For example, with an NGST detection limit, K$^{\prime}_{AB}$ =
31.4, we predict $\sim$ 8 SNe for the two low values of $\tau$, and
$\sim$ 5 for $\tau$ = 3 Gyr. Observations in the I band with limits
and field as for the VLT/FORS (see Table \ref{Table2}) results in 0.8,
1.1, and 1.8 SNe for increasing values of $\tau$. Counts may therefore
seem like a useful probe to distinguish between
different progenitor models (Ruiz-Lapuente \& Canal 1998). However,
the uncertainty in the modeling, especially in the normalization of
the Type Ia rates to the local value, makes the counts highly model
dependent. In next section we show that this is further hampered by
the additional dispersion introduced when considering alternative star
formation and extinction models.

If both SN type and redshift information are available for the
observed SNe, it may be possible to use the redshift distribution of
the SNe to distinguish between progenitor scenarios. Figs.  \ref{Fig1}
and \ref{Fig5} show that the peak in the SNR moves to lower $z$ as
$\tau$ increases. Also, the high redshift cutoff in the rates depends
strongly on $\tau$. Fig. \ref{Fig1} shows that the rates decrease
towards zero at redshifts $z \sim$ 5.5, $z \sim$ 3.5 and $z \sim$ 1.5,
for models with $\tau$ = 0.3 Gyr, $\tau$ = 1 Gyr and $\tau$ = 3 Gyr,
respectively. To reach these redshifts, filters unaffected by the UV
cutoff must be used (i.e., $\lambda_{eff} \gsim$ 4000(1+$z$)
\AA). Fig. \ref{Fig5} shows that the I filter is insensitive to Type
Ia's at $z\ \gsim$ 1.5, due to the spectral cutoff. Using the
K$^{\prime}$ filter (lower panels of Fig.  \ref{Fig5}) makes it
possible to sample all types of SNe up to $z\sim$ 5, which is
therefore most suitable for distinguishing different progenitor
models. As already mentioned, this requires a determination of both SN
type and redshift. A discussion of methods and problems regarding this
follows in Sect. 7. 

The only observational estimate of a Type Ia SNR at moderate redshift
is by Pain et al. (1996). From a careful analysis, using realistic
light curves and spectra, they find a Type Ia rate at $z\sim$ 0.4 of
34.4$^{+23.9}_{-16.2}$ SNe yr$^{-1}$ deg$^{-2}$ for 21.5 $<$ R$_{AB}<$
22.5. Using the same magnitude interval, and counting the SNe
exploding during one year, we find $5-24$ SNe yr$^{-1}$ deg$^{-2}$ for
$\tau$ = 0.3 - 3 Gyr, where the lower number corresponds to the $\tau$
= 0.3 Gyr model. The mean redshift of these SNe is in our calculation
$<z>\sim$ 0.35. These results seem to agree well, possibly favoring a
high value of $\tau$. Note, however, that our estimated rate of Type
Ia's is highly dependent on the normalization set by the local rate of
these SNe (i.e., the efficiency parameter $\eta$). We have already
seen that the uncertainty in this normalization may be a factor $\sim$
3. 
\subsubsection{Number of pre-maximum Type Ia SNe}
\begin{figure}
\resizebox{7.0cm}{!}{\includegraphics{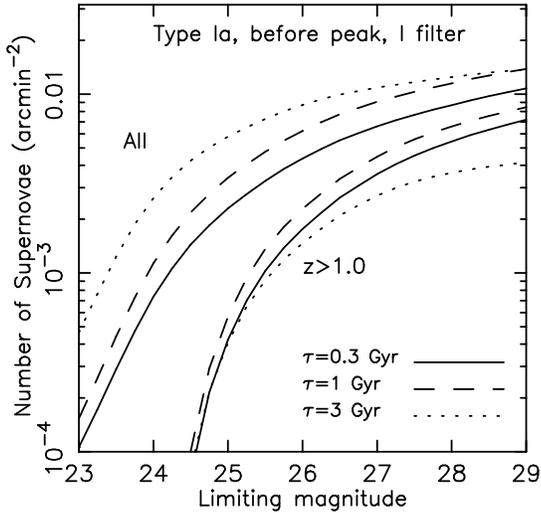}} \hfill \vskip1mm
\parbox[b]{8.5cm}{
\caption{The number of simultaneously detectable Type Ia SNe on the
rising part of the light curve for the I filter for three different
values of $\tau$. Upper sets of curves show the total number of SNe,
while the lower show SNe with $z>$1.}
\label{Fig6}}
\end{figure}

The number of simultaneously detectable SNe discussed above is a
result of events over the whole light curve. Using Type Ia's as
standard candles for determination of $\Omega_0$ requires observations
at the peak of the light curve, i.e that a first detection is made at
the rising part of the light curve. We estimate the number of such SNe
by assuming that the comoving rise time is 15 days. Fig. \ref{Fig6}
shows the number of Type Ia SNe down to different limiting magnitudes
before the peak of the light curve in the I filter for the three
values of $\tau$. Also shown is the number of such SNe with $z >$ 1
(the lower sets of curves). An I band survey covering a one square
degree field with limiting magnitude I$_{AB}\sim$ 27, will detect
$16-31$ pre-maximum Type Ia SNe (lower numbers for lower values of
$\tau$), corresponding to $\sim$ 30\% of the total number. About $5-8$
of the pre-maximum Type Ia's have redshifts $z>$ 1. These numbers are,
again, sensitive to the local rate of Type Ia SN.
\section{Alternative star formation scenarios}
\subsection{High extinction models}
There are several claims that an extinction $E_{B-V}$=0.1, as
estimated by Madau (1998) and used here, is too low. Meurer et
al. (1997) use the ratio between the far-IR and UV fluxes as a probe
of the dust extinction in a galaxy, and find that the high redshift UV
dropout galaxies may have a UV absorption of 2-3 mag. Using ISO
observations of the HDF, Rowan-Robinson et al. (1997) estimate that
only a third of the star formation is revealed by the UV-luminosity,
with the rest shrouded by dust. Observations with SCUBA at 850 $\mu$m
(Hughes et al. 1998; Smail et al. 1997) also indicate that a large
fraction of the early star formation is hidden by dust. Several other
estimates, based on observations of high-$z$ galaxies (eg. Sawicki \&
Yee 1998; Ellingson et al. 1996; Soifer et al. 1998), support a higher
extinction of $E_{B-V}\sim$ 0.3. Observations of the far-infrared
extragalactic background (Burigana et al. 1997) also seem consistent
with a higher extinction. 

With an extinction of $E_{B-V}\sim$ 0.3 it is likely that the observed
peak in the SFR derived from UV-luminosities is illusive, and that the
star formation history is compatible with a more constant rate, as in
a monolithic collapse scenario (e.g.,  Larson 1974; Ortolani et
al. 1995). In this model galaxy formation is thought to have occurred
during a relatively short epoch at high redshift, $z_F\ \gsim$ 5. This
yields a SFR that increases from $z$ = 0 to $z\sim$ 1.5 and then stays
almost constant up to $z\ \gsim$ 5. A SFR of this form may be
compatible also with a hierarchical model, as shown in a recent paper
by Somerville \& Primack (1998).

It should be noted that although there is much in favor for models
with a flat SFR to $z \gsim$ 5, they tend to over-predict the
metallicity in the region $z \simeq$ 2 - 3, compared to estimates from
damped Lyman-$\alpha$ systems (Blain et al. 1999; Madau et al. 1998b;
Pei et al. 1998). These models also seem to over-predict the local
K-band luminosity (Madau et al. 1998b).

We have calculated the SFR and the resulting SNRs for this high dust
scenario by adjusting the observed UV luminosities of Lilly et
al. (1996) and Madau (1998) to an extinction E$_{B-V}$=0.3. We also
increase the local SFR in this model so that the observed B band
luminosity density of Ellis et al. (1996) is reproduced. The onset of
star formation is set to $z_F$ = 5, while at even higher redshifts the
SFR declines quickly. In Sect. 4.3 we comment on how higher values of
$z_F$ affect the estimated rates. For the normalization of the type
Ia's at $z$ = 0, we assume that the increase in the intrinsic local
B-band luminosity density should lead to an increase in the Type Ia
rate by approximately the same amount. In Fig. \ref{Fig7} we show this
SFR and the intrinsic SNR for both core collapse SNe and Type Ia SNe,
assuming $\tau$ = 1 Gyr (dashed lines). 

\subsection{Low extinction models}
The model above combines a high star formation at high redshifts with
a high extinction in order to match the same observed luminosity
densities. Another alternative scenario is an increased SFR at high
$z$, but a low extinction, as in the original hierarchical
model. Evidence for such a scenario come from Pascarelle et
al. (1998), who calculate the evolution of the UV luminosity density
up to $z\sim$ 6 from $\sim$ 1000 galaxies with photometric redshifts
in the HDF. Taking the effects of cosmological surface brightness
dimming into account and using a limiting surface brightness
independent of redshift, they argue that the claimed UV luminosity
density decrease at $z>2$ is mainly caused by a selection
effect. Further, Hu et al. (1998) show that an increasing fraction of
the total SFR at high redshift takes place in Ly$\alpha$ emitters, and
find that the fraction at $z\sim$ 3 may be comparable to that derived
from the dropout galaxies. Most of these objects would not show up in
a survey such as the HDF, due to their low surface brightness. There
is, however, no reason why SNe in these galaxies should not be
detectable, especially since the authors argue that the extinction in
these objects should be low.

\begin{figure}
\resizebox{8cm}{!}{\includegraphics{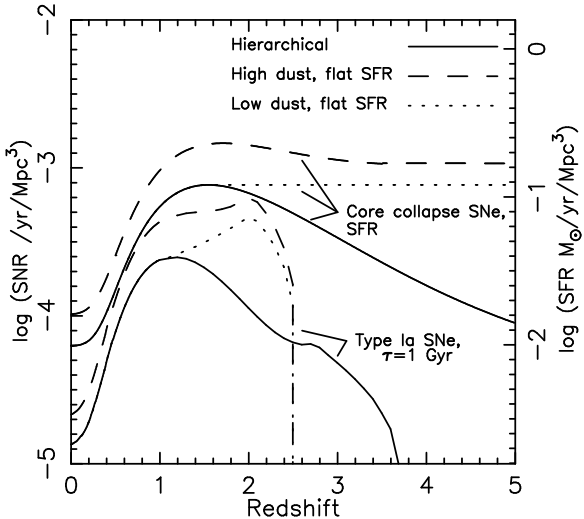}} \hfill \vskip1mm
\parbox[b]{8.5cm}{
\caption{Intrinsic SNRs for the three different models. Solid lines
show the hierarchical model with $E_{B-V}$ = 0.1, dashed lines show
the high dust model with $E_{B-V}$ = 0.3 and flat SFR at high $z$, and
dotted lines show the model with $E_{B-V}$ = 0.1 and a flat SFR at
high $z$. Upper lines show the rates of core collapse SNe. These scale
directly with the SFR, given on the right axis. Lower lines show the
rate of Type Ia, for $\tau$ = 1.0 Gyr model for each of the SFR
models.}
\label{Fig7}}
\end{figure}

To mimic this scenario we have constructed a SFR with low extinction,
E$_{B-V}$=0.1, that increases like the hierarchical model up to
$z\sim$ 1.5, but stays flat at higher redshifts up to a formation
redshift, $z_F$ = 5. The normalization of the Type Ia's at $z$ = 0 is
the same as in the hierarchical model. This SFR and corresponding SNRs
are shown in Fig. \ref{Fig7} as the dotted line.
\subsection{Results for alternative dust and star formation models}
We have repeated our calculations in Sect. 3 for the two additional
scenarios described above. In the optical filters (probing rest-frame
UV to optical), the increased SFR in the high dust model is mostly
compensated for by higher extinction, resulting in observed rates that
are only $\sim$ 10\% above those in the hierarchical scenario. The NIR
filters (probing rest-frame optical to NIR) are less affected by
extinction, which leads to a factor $\sim$ 2 higher rates in the high
dust model, compared to the hierarchical model. The estimates differ
even more when comparing high redshift subsamples, i.e. when $z\
\gsim$ 2 is observed. This illustrates that the observed SNR may 
serve as an
independent probe for the instantaneous SFR that is not subject to the
same high uncertainty due to the unknown amount of dust extinction as
the UV-luminosity is.

For comparatively bright limiting magnitudes, that do not probe SNe
above the proposed peak in the hierarchical scenario, the two models
with $E_{B-V}$=0.1 give, by construction, the same
result. Observations must reach SNe at $z \sim$ 2 before the rates
start to differ. 

Except for the rare Type IIn's, a SN at peak magnitude at $z \sim$ 2
has K$^{\prime}_{AB} \gsim$ 27, making NGST necessary for this type of
observations. As an illustration, Table \ref{table3} shows to what
extent rates of core collapse SNe are useful to constrain the SFR at
high redshifts. For core collapse SNe with $z\ \gsim$ 2, NGST should
detect $\sim$ 3 and $\sim$ 2 times higher rates for the
high-dust-flat-SFR model and the low-dust-flat-SFR model,
respectively, compared to the hierarchical model. At $z \gsim$ 4 these
factors are $\sim$ 5 and $\sim$ 4.
\begin{figure*}
\resizebox{3.5cm}{!}{\includegraphics{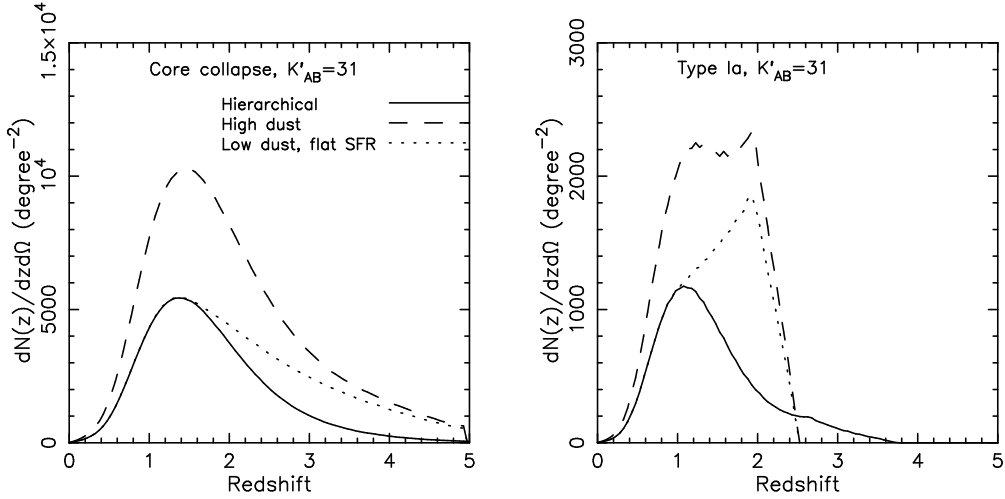}} \hfill \vskip1mm
\parbox[b]{17cm}{
\caption{Redshift distribution for core collapse and Type Ia SNe
observed to K$^{\prime}_{AB}$=31 for the three different SFR
models. Results for core collapse SNe are shown in the left panel,
while the right panel shows the results for Type Ia SNe with
$\tau$=1.0 Gyr. Note the different scales on the y-axis.}
\label{Fig8}}
\end{figure*}
\begin{table*}
\begin{center}
\begin{tabular}{|l|c|c|ccccc|cc|}
\hline Model& $E_{B-V}$ & Limit & N(tot) & N($z>$1) & N($z>$2) &
N($z>$4) & N($z>$5)& N($z>$5)& N($z>$5) \\ &   & m$_{AB}$ &  &  &  & &
& $z_F$ = 7  &  $z_F$ = 10 \\ \hline Hierarchical & 0.1 &
M$^{\prime}$=30 & 2.2 & 1.9 & 0.85 & 0.11 & 0.05& 0.05& 0.05\\ &  &
K$^{\prime}$=31 & 2.6 & 2.2 & 0.86 & 0.06 & 0.02& 0.02& 0.02\\ \hline
High dust, & 0.3 & M$^{\prime}$=30 & 6.6 & 6.0 & 3.6 & 0.72 & - & 0.75
& 1.0\\ flat SFR &  & K$^{\prime}$=31 & 5.8 & 5.2 & 2.6 & 0.29 & - &
0.15 & 0.20\\ \hline Low dust,  & 0.1 & M$^{\prime}$=30 & 3.2 & 2.9 &
1.9 & 0.42 & - & 0.44 & 0.63\\ flat SFR &  & K$^{\prime}$=31 & 3.5 &
3.1 & 1.7 & 0.25 & - & 0.14 & 0.20\\ \hline
\end{tabular}
\end{center}
\caption{Estimated number of core collapse SNe per sq. arcmin in the
M$^{\prime}$ and K$^{\prime}$ band for different SFR and dust
models. N(tot) gives the total number of SNe, while the other columns
gives the number of SNe with a redshift above different specified
values. The two models with flat SFR at high $z$ has a formation
redshift $z_F$ = 5, except in the two rightmost columns where the
formation redshift is set to $z_F$ = 7 and $z_F$ = 10, respectively.}
\label{table3}
\end{table*}

Due to the short time interval between formation and explosion in the
case of core collapse SNe the choice of formation redshift enters only
for observations that are deep enough to actually probe $z_F$. Table
\ref{table3} shows the variation of the estimated number of core
collapse SNe with $z >$ 5 in the two models with flat SFR as $z_F$ is
changed from  $z_F$ = 5 to $z_F$ = 7 and $z_F$ = 10. Note that more
than 10\% of the SNe in the M$^{\prime}$ band have a redshift $\gsim$
5 in these models, compared to $\sim$ 2\% in the hierarchical model.

Our calculations show a modest increase in the number of Type Ia's for
the high dust scenario, compared to the hierarchical scenario, at
magnitudes brighter than $\sim$ 25. At magnitudes corresponding to $z
\gsim$ 1, the differences are considerably larger. This is also true
for the scenario with flat high-$z$ SFR and low extinction, but the
increase is modest, and it does not start before $z \sim$ 1.5 As an
illustration, Fig. \ref{Fig8} shows for the $\tau$ = 1 Gyr model that 
for K$^{\prime}_{AB} \lsim$ 31.4, the number of Type Ia SNe increases by 
a factor $\sim$ 2 for the two
models with flat SFR at high $z$, compared to the hierarchical model. 
Including the full range of $\tau$
and different star formation scenarios, the NGST should detect between
5-25 Type Ia SNe per field in the K$^{\prime}$ band. For the ground
based limit, I$_{AB}$ = 27, the number of Type Ia SNe increases by a
factor 1.4-1.9 for the high dust scenario, and a factor 1.0-1.8 for
the low dust scenario with flat SFR (smaller increase for lower values
of $\tau$).

If $\tau$ is long, stars formed at early cosmological epochs may
survive until low redshifts before ending as Type Ia SNe. Therefore,
the increased star formation at high redshifts in the two constant SFR
models makes the rate of Type Ia SNe sensitive to the assumed
formation redshift, $z_F$. The choice of $z_F$ affects the SNR down to
a redshift $z$, corresponding to a time $t = t_F + \Delta t_{MS} +
\tau$. Increasing $z_F$ to $>$ 5 increases the rates at $z\gsim$ 2.7,
$z\gsim$ 2.0 and $z\gsim$ 1.0 for $\tau$ = 0.3, 1 and 3 Gyr,
respectively. For example, with $\tau$ = 1 Gyr the high-$z$ cutoff in
the Type Ia rate occurs at $z \sim$ 2.5 for the two models with flat
SFR and $z_F$ = 5 (see Fig. \ref{Fig7}). With $z_F$ = 7 and $z_F$ = 10
the cutoff in the rates moves to $z \sim$ 3.0 and $z \sim$ 3.4,
respectively. This effect is larger for smaller values of $\tau$.

As earlier mentioned, a further possibility that may increase the
predicted counts of core collapse SNe is if the first SNe and their
progenitors sweep away the dust, making the extinction lower for a
large fraction of the SNe relative to the stars that produce the UV
luminosity. We have studied this scenario by simply neglecting the
extinction of the SNe in the two models with $E_{B-V}$ = 0.1 (not
included in Table \ref{table3}). The main difference occurs for the
optical bands, where the effect of a lower extinction is largest. In
the I band the counts increase by a factor $\sim$ 2, whereas the
counts increase by a factor $\sim$ 1.4 in the K$^{\prime}$ band for
limits K$^{\prime}_{AB} \lsim$ 27. At fainter K$^{\prime}$ limits,
i.e. reaching beyond the peak region, the differences decrease because
a majority of all SNe is detected in both models, despite the
different amount of absorption. For K$^{\prime}_{AB}$ = 31.4 the
increase is therefore only $\sim$ 10\%.
\section{Dependence on cosmology}
Besides the standard flat $\Omega_M$ = 1 cosmology (SCDM) used so far, we have also studied two additional cosmologies; one open cosmology (OCDM) with $\Omega_M$ = 0.3 and $\Omega_{\Lambda}$ = 0, and one flat, $\Lambda$-dominated cosmology ($\Lambda$CDM) with $\Omega_M$ = 0.3 and $\Omega_{\Lambda}$ = 0.7. The most obvious effect of a different cosmology comes from the luminosity distance, $d_L$ (Eq. 5). Changing from SCDM to OCDM and $\Lambda$CDM increases the distance modulus, making the high-$z$ SNe fainter (Eqs. 4 \& 5), as seen in Fig. \ref{Fig2}. The increased distance modulus in the OCDM and $\Lambda$CDM cosmologies also affects the intrinsic star formation rates used. A larger distance modulus implies an increase in the absolute magnitudes of the galaxies from which the star formation rate, and hence also the supernova rates, are derived. As expected, these two effects, i.e fainter apparent SNe and an increased SNR, almost cancel when it comes to the observed number of core collapse SNe per square angle.\

The cosmology also affects the volume element, given by
\begin{equation}
dV=\frac{d^2_M}{(1+\Omega_kH_0^2d^2_M)^{1/2}} d(d_M)d\Omega,
\end{equation}
where $d_M$ is the proper motion distance, $d_M$ = (1+$z)^{-1}d_L$. The luminosity distance, $d_L$, is given by Eq. (5). The change in volume element affects the SNRs when expressed in units number of SNe per Mpc$^3$ per yr. It does not, however, change the estimated rates of observed core collapse SNe expressed in number of SNe per square angle. This is due to the fact that the core collapse SNR is directly proportional the observed luminosity density of galaxies, which is calculated from the number of galaxies per redshift interval for a specific angle over the sky, an observational quantity independent of cosmology.
 
When it comes to the Type Ia SNe, the volume element enters the calculations since the time between the formation of the progenitor star and the explosion of the SN may be a significant fraction of the Hubble time. This dependency on cosmology increases as the delay time $\tau$ increases. The general trend is an increased SNR for the alternative cosmologies at high redshift. 

\begin{figure*}
\resizebox{3.5cm}{!}{\includegraphics{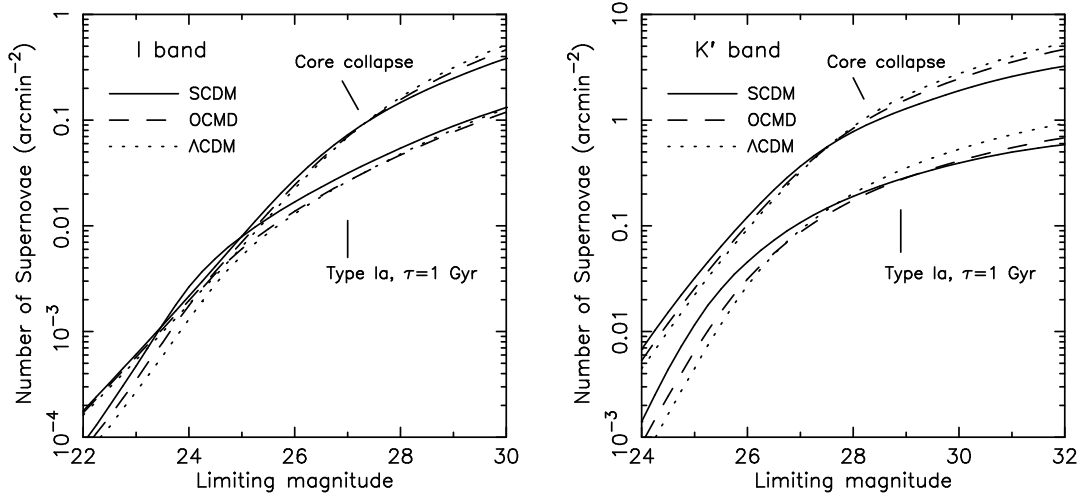}} 
\hfill 
\vskip1mm
\parbox[b]{17cm}{
\caption{Number of SNe per square arcmin that can be detected down to
different limiting magnitudes in the I and K$^{\prime}$ bands for three different cosmologies; SCDM ($\Omega_M$ = 1, $\Omega_{\Lambda}$ = 0), OCDM ($\Omega_M$ = 0.3, $\Omega_{\Lambda}$ = 0) and $\Lambda$CDM ($\Omega_M$ = 0.3, $\Omega_{\Lambda}$ = 0.7). For the Type Ia rates we assume a delay time $\tau$ = 1 Gyr.}
\label{Fig14}}
\end{figure*}

Fig. \ref{Fig14} show the estimated number per square arcmin of core collapse and Type Ia (using $\tau$ = 1 Gyr) SNe in the I and K$^{\prime}$ filters down to different limiting magnitudes for the three cosmologies, using the hierarchical model for star formation.
For the number of core collapse SNe the effect of changing
 cosmology is small for the reasons discussed above. Only at the very faintest magnitudes is there significant deviation between the models. At K$^{\prime}$ = 31, the estimated rates increase by a factor $\sim$ 1.5  when changing from SCDM to $\Lambda$CDM.

The estimated rates of Type Ia SNe differ at faint magnitudes by up to a factor two between the cosmologies. For $\tau$ = 1 Gyr the intrinsic rates of the OCDM and $\Lambda$CDM models are higher than the SCDM at redshifts $z \gsim$ 1. It is, however, necessary to reach faint magnitudes (K$^{\prime}\sim$ 28) to observe this increase in the total rates. Note, however, that a Type Ia at peak magnitude has m $\sim$ 25 at $z \sim$ 1. This means that using SNe seen at peak allows probing of the region where the rates start to differ between the cosmological models at more moderate limiting magnitudes. Ruiz-Lapuente \& Canal (1998) discuss the use of Type Ia's to distinguish between different cosmologies. In Sect. 8 we comment on their results.

The increased Type Ia rates at high $z$ for the alternative cosmologies means that the redshift cutoff moves to higher redshifts. This cutoff is also highly dependent on the delay time $\tau$ (see Fig. \ref{Fig1}). It is, however, somewhat less dependent on the SFR. Therefore, if $\tau$ is known, it should be possible to constrain the cosmology, even if the SFR is not accurately known. As an illustration, including all SFR scenarios, and using $\tau$ = 1 Gyr ($\tau$ = 3 Gyr), results in a cutoff in the Type Ia rates at $z$ 2.5 - 3.5 (1.2 - 1.5) for the SCDM cosmology, while the drop is at $z$ 2.9- 4.9 (1.6 - 2.2) and $z$ 3.2 - 5.3 (1.9 - 2.6) for the OCDM and $\Lambda$CDM cosmologies, respectively. The higher values in these redshift ranges correspond to the hierarchical star formation scenario.
\section{Gravitational lensing}
In a recent paper Marri \& Ferrara (1998) study the effects of
gravitational lensing (GL) on high-$z$ objects, in particular Type II
SNe, for a hierarchical model of galaxy formation. For the three flat 
cosmologies
they study (SCDM with $\Omega_M$ = 1, LCDM with $\Omega_M$ = 0.4, 
$\Omega_{\Lambda}$ = 0.6 and CHDM with $\Omega_M$ = 0.7, 
$\Omega_{\nu}$ = 0.3), they find that there is at least a
10\% chance that objects with $z\ge$ 4 are magnified by a factor
$\gsim$ 3. To estimate the effects of GL on our results we have
therefore used their magnification probabilities on our model 
(i.e. their Figs. 4 and 5). 

For the SCDM model, which yields the highest magnification, we find
that the increase in the total number of core collapse SNe is only a
few percent when using the NGST limits (M$^{\prime}_{AB}$ = 30.2 and
K$^{\prime}_{AB}$ = 31.4). The effects are even smaller for the shorter
wavelength bands. The reason for the small effect is that with the
faint limits of the NGST almost all SNe up to $z\sim$ 4 are detected
even without magnification, and that the number of SNe with even
higher $z$, where the magnification has largest effect, is relatively
small. 

More interesting, for SNe with $z \ge$4 we find, when using the NGST
M$^{\prime}$ limit, an increase by $\sim$ 20\% of the predicted
counts. For SNe with $z \gsim$ 9 the increase is $\sim$ 40\%. It
should, however, be noted that estimates presented by Porciani \&
Madau (1998) give a much lower probability for substantial
magnification compared to the Marri \& Ferrara results used in the
estimates above. The reason for this seems to be that Marri \& Ferrara
assume point-like lenses, whereas Porciani \& Madau use a more
realistic mass distribution characterized by singular isothermal
spheres. Therefore, the effects of gravitational lensing presented
here could be overestimated. 
\section{Redshift determinations}
A major problem when comparing predictions from different scenarios to
observations is the determination of the SN redshift. A direct
spectroscopic determination with a resolution of $\gsim$ 100 is only
possible for SNe more than two magnitudes brighter that the limiting
magnitude, i.e. I$_{AB}\sim$ 25 for $8-10$ m class telescopes and
K$^{\prime}_{AB}\sim$ 28.5 for the NGST. To reach fainter magnitudes
the  main alternative is photometric redshifts of either the host
galaxy or  the SN. Photometric redshifts for galaxies have been
discussed  extensively by e.g., Fern\'{a}ndez-Soto
et al. (1999), Yee et al. (1998), Gwyn (1995) and Connolly et al. (1995). 
The fact that
a large fraction of the star formation up to $z \gsim$ 1 occurs in
dwarf galaxies, as well as the cosmological dimming, can make such a
determination difficult. An alternative is to estimate a photometric
redshift directly from the SN. A problem here is that the SN spectrum
changes with both type and epoch.

\begin{figure*}
\resizebox{3.5cm}{!}{\includegraphics{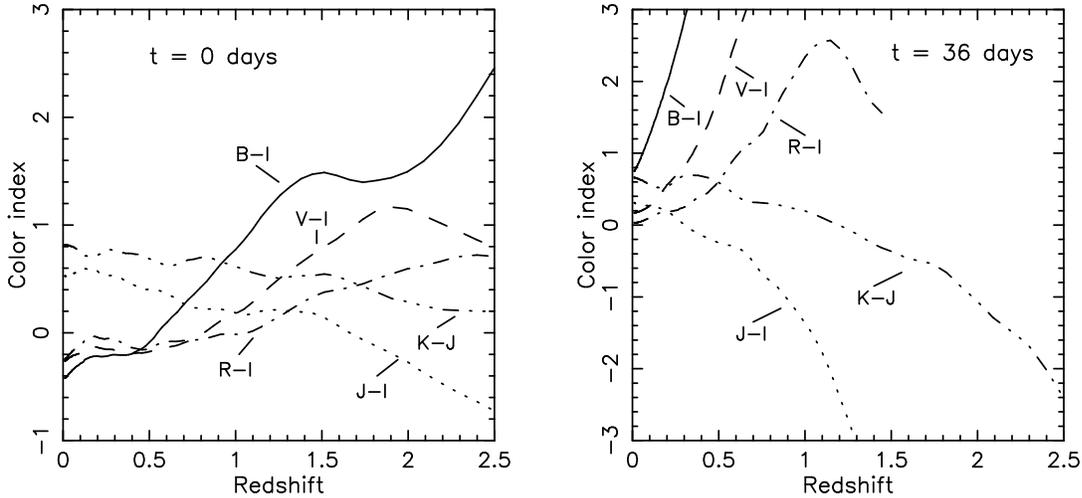}} \hfill \vskip1mm
\parbox[b]{17cm}{
\caption{Color indices for Type IIP SNe. The left panel shows indices
at the peak, while the right panel shows indices at the plateau phase
$\sim$ 36 days after the peak.}
\label{Fig10}}
\end{figure*}

To examine this possibility we have determined broad band colors for
the different types of SNe as function of epoch and redshift. As an
example we show in Fig. \ref{Fig10} the color indices for a Type IIP
SN as function of redshift at different phases. The spectra are taken
from the synthetic spectra calculated by Eastman et al. (1994). These
spectra assume LTE, but may nevertheless give a good impression of the
qualitative evolution. 

At early time, before $\sim$ 20 days the spectrum is a fairly smooth
blackbody without a strong UV cutoff. The color indices do therefore
not change dramatically with redshift. At later stages in the  plateau
phase the spectrum does not change much. An important aspect  is that
the UV cutoff at $\sim$ 4000 \AA\ has now developed to its  full
extent, and the UV is essentially black. This is probably the  most
useful feature for identifying high redshift SNe  photometrically. The
extent of this UV drop may, as we discuss below,  depend on the
metallicity. The UV cutoff has a very pronounced effect  on the
optical color indices at $z\ \lsim$ 1, with strong increases in  the
B-I, V-I and R-I indices at successively larger $z$. For $z\ \gsim$ 1
the J-I and K-I, and finally K-J, are most useful due to the essential
disappearance of the SNe in the optical. 

A problem with photometric redshifts of SNe, compared to galaxies, is
that the colors, as we have seen, depend sensitively on the epoch. In
addition, they depend on SN type. E.g., Type IIL's have less UV
blanketing, while the Type Ia's have a rapid development of the UV
cutoff. To break this strong degeneracy it is essential to have
information about both SN type and epoch. It is therefore necessary to
obtain reasonable light curves, i.e. a fairly large number, $\gsim$
5-10, observations of the field. A complete analysis can therefore be
quite costly in terms of observing time. 

An alternative redshift method may be to use reasonably well sampled
light curves in combination with the cosmic time dilation. For Type Ia
SNe one can safely assume a standard light curve. Although the
absolute luminosity can vary by a large factor, Type IIP's have a
fairly well defined duration of the plateau phase, which lasts $\sim$
100 days. Also Type IIL's and Ib's and Ic's have reasonably
standardized light curves. From the observed light curve one can then,
at least for SNe with $z \gsim$ 1, get an approximate redshift within
$\sim$ 25\% from a comparison with the low $z$ light curve
templates. However, since the light curve should be followed over a
decline of about $\sim$ 2 magnitudes in order to achieve a type
specific light curve, the gain in depth by using photometry instead of
multi object spectroscopy is marginal. The photometric accuracy also decreases for these levels and the actual limit may be even higher than two magnitudes. Therefore, in practice little is probably gained compared to direct spectroscopy.
\section{Discussion}
\subsection{Uncertainties in the models}
A major source of uncertainty is the treatment of dust extinction. In Sect. 4 we showed how different assumptions about the dust extinction affect the estimated rates. An underlying assumption in each of the calculations is that the same amount of dust affects the UV-luminosity from the high-$z$ galaxies (used to calculate the SNR) and the light from the SNe. If the
UV-luminosity originates from regions with dust properties that differ
considerably from the regions where the SNe originates, an extra dispersion in the estimated rates should be expected. It is, however,
difficult to estimate the uncertainties in the rates introduced by this,
since the distribution of the dust within the galaxies at high $z$ is
poorly known.

Other sources of uncertainties are the range
of progenitor masses in Eq. (2) and the choice of IMF (Sect. 2.2), as well as the dependence on cosmology (Sect. 5). Also, the
distribution of the different types among the core collapse SNe influences the estimates. Changes in the fractions of the faint
SN1987A-like or bright Type IIn's are most important. Due to their low
luminosity, the SN1987A-like SNe will be too faint to be detected by
ground based telescopes for $z\gsim$ 0.5 (I$_{AB}\gsim$ 27 at
peak). Doubling the fraction of these SNe from 15 \% to 30 \%, or, at
the other extreme, setting the fraction to zero, changes the total
number of detected SNe by $\pm$18\% at these redshifts. Changing the
fraction of Type IIn's has a slightly different effect. By increasing
$f$(IIn) from 2\% to 4\%, which is the upper limit proposed by
Cappellaro et al. (1997), the total number of detected core collapse
SNe at I$_{AB}\sim$ 24 increases by $\sim$ 17\%. At these magnitudes
it is mainly low redshift SNe in the steep part of the SFR curve that
contribute, explaining the relative large effect. At fainter limiting
magnitudes, an increasing fraction of normal luminosity SNe are
detected at, and even beyond, the peak. The relative increase due to
the Type IIn's is therefore marginal. For I$_{AB}$ = 27 the number of
SNe increases by $\sim$ 4\% and for K$^{\prime}_{AB}\sim$ 31.4 the
number increases by $\sim$ 1\%. 

Adding the uncertainties, we find that the counts of core
collapse SNe may vary by a factor more than two due to insufficiently
known model parameters.

The estimated rates of Type Ia SNe are subjected to even larger
uncertainties. Besides the factors which also affect the core collapse
SNe, the rates of Type Ia's depend strongly on the assumed progenitor
scenario, and are more dependent on cosmology. Also, the normalization at $z$ = 0
introduces an additional uncertainty by a factor $\sim$ 3. Considering
this, it seems unlikely that counts of Type Ia's can be used to set
any constraints on the model parameters. With additional information
about the redshift distribution of the SNe it should, however, be
possible to constrain either the progenitor life time or the cosmology. One
of these parameters must, however, be independently determined, since
there is a degeneracy between the both when it comes to the estimated
redshift distribution of the SNe. To decrease the uncertainties
involved, well defined searches of SNe at low redshift are highly
desirable.
\subsection{Dependence on metallicity}
With increasing redshift the mean metallicity decreases, although the
dispersion may be higher that at present. It is therefore interesting
to discuss the consequences of a lower general metallicity. 

For Type Ia SNe, Kobayashi et al. (1998) argue that for one of the
most likely progenitor scenarios, based on super-soft X-ray binaries,
the necessary condition for a Chandrasekhar mass explosion may not
occur for a metallicity of $Z\ \lsim$ 0.1 $Z_{\odot}$, consistent with
the decrease in the galactic [Fe/O] ratio at this metallicity. The
physical reason can be traced to the peak in the opacity curve at
$\sim$ 10$^5$ K, produced by iron. In the scenario by Kobayashi et
al., this corresponds to a drop in the Type Ia rate at $z\sim$ 1.2. At
higher redshifts Type Ia explosions are inhibited, since [Fe/H] $\lsim$
- 1 here. Including the possibility of a dispersion in the evolution
of the metallicity leads Kobayashi et al. to conclude that a cutoff in
Type Ia rates should occur at $z$ = $1-2$. The Kobayashi et
al. scenario should observationally be similar to the $\tau$ = 3 Gyr
model, with a turn-on at $z \sim$ 1.5. Because the life time in their
model is $\tau$ = 0.6 Gyr, the peak in the SNR should occur at $z
\gsim$ 1, rather than at $z \sim$ 0.7, as in the $\tau$ = 3 Gyr model.

For core collapse SNe a lower metallicity can have several
effects. First, line blanketing in the UV may decrease somewhat for
the Type IIP's. Since this is mainly an ionization effect, rather than
an abundance effect, it is, however, likely that this effect is
small. This is partly confirmed by the calculations by Eastman et
al. (1994), who find only a marginal decrease of the UV blanketing as
the metallicity is decreased from solar to a tenth of solar. For
extremely low metallicity, like for Pop. III stars with $Z\ \lsim$
10$^{-2}Z_{\odot}$, blanketing may, however, decrease significantly,
although the Balmer jump will still be present.

Secondly, the relative fraction between different SN types may
change. In particular, the number of blue, compact supergiant
progenitors similar to Sanduleak -69$^{\circ}$ 202 may increase. This
would then lead to a larger fraction of faint SN1987A like SNe,
decreasing the number of observable core collapse SNe at high $z$. We
emphasize that the exact reason for the blue progenitor of SN1987A is
not fully understood (e.g., Woosley et al. 1999).

Finally, the mass loss process of the SN progenitor may depend on the
metallicity. An example is the known decrease of the mass loss rate
with  decreasing metallicity for radiatively accelerated winds (e.g.,
Kudritzki et al. 1987). In the red supergiant phase dust driven mass
loss may be less efficient (Salasnich et al. 1999). The importance of
binary mass transfer may also depend on the metallicity. A change in
the mass loss rate with metallicity would change the relative
proportions between the different core collapse types. In particular,
a decrease of the total mass lost is expected to lead to a decrease in
the number  of Type IIL, Ib and Ic SNe, while favoring Type IIP's. In
addition, a less  dense circumstellar medium medium could then lead to
a decrease in the  ionization by the circumstellar interaction, and
stronger line blanketing for  the Type IIL and IIn's. Note, however,
that the mass loss process even for local red supergiants in their
final phase is poorly understood.
\subsection{SNe as probes of star formation}
With large ground-based telescopes, and especially with NGST, it
should be possible to detect SNe up to high redshifts, and to estimate
the rates of both core collapse and Type Ia SNe. We have shown that
because the rate of core collapse SNe follows the SFR, it should be
possible to use observed rates of these SNe to constrain the SFR. As
we have seen, a major problem is the influence of dust extinction. In
this respect we note that the NIR bands have the advantage of being
less affected by dust extinction than the observed UV-luminosities. At
high redshifts these bands correspond to the optical rest wavelength
bands, and have therefore a factor of $2-3$ lower extinction than the
UV bands. Different star formation models may therefore better be
tested by using the K and M bands. Reaching high redshifts ($z\ \gsim$
2), corresponding to K$^\prime_{AB}\ \gsim$ 27, increases the
differences in the predicted counts between the star formation models
significantly (see Sect. 4.3).

The estimated difference in the redshift-integrated counts between the
hierarchical star formation model and the high dust star formation
model is a factor $\gsim$ 2. This is about the same factor produced by
the uncertainties in the modeling. It is therefore difficult to use
these counts to probe star formation scenarios, unless the parameters
involved are better known. What seems more feasible is to use redshift
subsamples to constrain the shape of the SFR. As shown in Sect. 4.3, for
models with a flat SFR at high $z$, we estimate a factor $\sim$ $4-5$
more SNe with $z\ \gsim$ 4. A major problem here is to determine the
redshifts of the SNe (see discussion in Sect. 7).

A further problem is if a large fraction of the star formation takes
place in galaxies with very large extinction, like M 82 or Abell 220
with $A_V \sim 5-10$ magnitudes (instead of $A_V\ \sim\ 1$, as for our
high dust model). The difference between the optical and UV extinction
is then of less importance, leading to a decrease in the estimated
differences between the models. A similar large extinction may be
indicated by the results of the ISO observations of some deep fields
(e.g., Flores et. al. 1999). Far-IR observations is then the most
reliable way of deriving the true star formation rate. An alternative
is to use some other source tightly coupled to star formation, but not
affected by dust absorption. If gamma-ray bursts are related to some
class of core collapse SNe (e.g., Type Ic's), they may be such class
of objects (Cen 1998). Also radio observations may be interesting in
this respect.
\subsection{SNe and nucleosynthesis}
The study of the nucleosynthesis by direct observation of SNe is
naturally affected by the same problem as the star formation
rate. Unless the dust extinction can be determined reliably in an
independent manner, the true number of SNe is difficult to derive. In
addition to this, the metallicity yields for SNe of different masses
is non-trivial to derive even at low redshifts (e.g., Fransson \&
Kozma 1999). Only for SN 1987A and a couple of other SNe has this
become possible. The alternative to use theoretical yields from
collapse calculations is obviously less satisfactory. The lower
metallicity may also affect e.g. the mass loss processes, as discussed
in Sect. 8.2. This may change both the progenitor structure and the upper
and lower limits for the core collapse and Type Ia SNe, as well as the
heavy element yields. In our view one of the most interesting goals
for the observation of SNe at high redshift may be to
observationally study the differences between the SNe in the early
universe and those today.
\subsection{Spacing between observations}
An important aspect concerning the detection of SNe is the spacing in
time between the observations. In order to detect the SN, the
magnitude has to change appreciably. The interval is primarily
dependent on the shape of the light curve. Near the peak, where the SN
changes relatively fast, a comparatively short time is
sufficient. This applies to searches where detection of SNe on the
rising part of the light curve is the main objective (e.g., searches
for Type Ia's for $\Omega_0$). Core collapse SNe, which have a flatter
decline of the light curve, need a longer spacing. This is especially
true for the Type IIP's, which in the plateau phase decline 
by $\lsim$ 1 mag. Unless a SN can be detected (against the host
galaxy) with this precision it will be missed. The limiting magnitude
of the search also affects the necessary spacing. A deeper search
results in a higher mean $z$ of the observed SNe. Due to the cosmic
time dilatation, the light curves of these SNe are stretched in time,
implying that a longer interval between two observations is needed,
$\sim$ 100(1 + $z$) days. Therefore, in order to detect these SNe a
deep observation with the VLT requires an interval of $\gsim$ 100
days, while a corresponding NGST observation requires approximately a
years interval.

\subsection{Comparison to other works}
Apart from the studies by Pain et al. (1996), Madau et al. (1998a) and
Chugai et al. (1999), which we have already commented on, there is a
number of related investigations.

Marri \& Ferrara (1998) have studied of the effects of gravitational
lensing of high redshift SNe. Using a Press-Schechter formalism and
gravitational ray-tracing, they determine the magnification
probability as function of redshift for different cosmologies. We have
already discussed the implications of their lensing results for our
simulations in Sect. 6. Marri \& Ferrara use these magnification
probabilities to estimate the observed magnitudes at high
redshift. The fact that there is a relatively large probability,
$\gsim$ 10\% for a factor of three or larger magnification for $z\
\gsim$ 4, means that even SNe as distant as $z \sim$ 10 may be within
the limits of NGST. When estimating the observed magnitudes they,
however, assume that the light curve is described by a Type IIP light
curve without any dispersion in magnitude, although as we have seen,
the Type IIP's show a very large variation in luminosity. They also
assume a fairly high temperature, $\sim$ 25\,000 K during the first 15
days, which is twice as high as the models by Eastman et al. (1994)
give. This is especially important for the high-$z$ SNe, and, as Marri
\& Ferrara show, a lower temperature makes the detectability
considerably more difficult. Marri \& Ferrara do not attempt any
discussion of expected rates of the high-$z$ SNe.

The effects of gravitational lensing is also investigated by Porciani
\& Madau (1998). They find, as earlier mentioned, a considerably lower
probability for a substantial magnification than Marri \& Ferrara
do. Porciani \& Madau present I band counts for Type Ia and core
collapse SNe, both including GL, and without lensing. These counts are
presented as the number of SNe in different magnitude bins (21 $<$
I$_{AB} <$ 27), seen at the peak of the light curve for an effective
observation duration of one year. This leads to lower estimates for
the observable number of SNe compared to our estimates, where we
include SNe detected over the whole light curve. To compare our
results we have calculated counts in the same units as used by
Porciani \& Madau. We find a fairly good  agreement between the core
collapse counts (deviation by a factor $\sim$ 2), but a somewhat worse
agreement between the Type Ia counts. It should be noted that
expressing rates in units of an effective observation duration
requires an idealized observational procedure (as we have shown in the
examples in Sect. 3.1).

Ruiz-Lapeunte \& Canal (1998) discuss the possibility of using R band
counts of Type Ia SNe to distinguish different progenitor
scenarios. They find, similar to our estimates, that models with
long-lived progenitors result in higher counts than models with
short-lived progenitors. To use this as a probe they note that it is
necessary to know the SFR better than a factor 1.5. However, the
uncertainty in the SFR seems, as we have shown, to be larger than
this. It should therefore be difficult to use counts to determine
progenitor scenarios. Additional information about the redshift
distribution of the SNe is required.

The same authors also estimate the effects on the counts for
alternative cosmologies. They find that a flat $\Lambda$-dominated 
universe ($\Lambda$CDM)
should result in higher counts of Type Ia SNe than a standard cold
dark matter universe (SCDM). The difference between the cosmologies
 start at m$_R \sim$ 24, and
increases at fainter limiting magnitudes. A somewhat
smaller increase in the counts is found for an open universe with zero
cosmological constant (OCDM).

Our results for different cosmologies agree with the general
trend of Ruiz-Lapeunte \& Canal. Using counts to distinguish between
cosmologies, however, requires both that the SFR is well known, and
that restrictions can be set on the progenitor life time. If this is
not the case, the degeneracy between the different parameters involved
makes a distinction between cosmologies very difficult.

In an interesting paper Miralda-Escud\'{e} \& Rees (1997) discuss the
possible detection of very high redshift core collapse SNe at $z\
\gsim$ 5. By requiring that a metallicity $\sim 10^{-2} Z_{\odot}$ is
produced at $z\sim$ 5, they estimate a rate of about one core collapse
SN per square arcmin per year above $z\sim$ 5. Our extrapolated
hierarchical model gives a rate of $\sim$ 0.05 SN per square arcmin
per year above $z\sim$ 5. This may favor models with a flat SFR at
high $z$, which result in $\sim$ 0.4 (0.7) SNe per square arcmin per
year above $z \sim$ 5 when using $z_F$ = 7 ($z_F$ = 10). However, the
metallicity used by Miralda-Escud\'{e} \& Rees may be overestimated by
an order of magnitude (Songaila 1997), leading to an overestimate of
the SNR by the same amount. Also, the redshift before which the
metallicity is assumed to have been produced affects the
comparison. Using $z\sim$ 3, instead of $z\sim$ 5, decreases the
estimated number SNe given by Miralda-Escud\'{e} \& Rees by $\sim$
30\%. More important, integrating our rates for redshifts above $z$ =
3, instead of $z$ = 5 as done above, results in a number of SNe that
is a factor $\sim$ 4 higher. Other uncertainties in the estimate by
Miralda-Escud\'{e} \& Rees include the actual fraction of the baryonic
matter which is enriched by the SNe.

Miralda-Escud\'{e} \& Rees limit their discussion to Type IIP SNe, and
do not attempt a detailed discussion of the observed rates. The
observed magnitudes compare fairly well with our magnitudes in the K
and M bands, but are brighter in the optical and near-IR bands. The
main reasons for this is that they use a higher effective temperature
and that they do not take into account any line blanketing in the UV,
as our models do. As we discuss in next section, the low metallicity
may decrease this effect. Apart from these caveats, the discussion by
Miralda-Escud\'{e} \& Rees provides an important constraint at high
redshifts.

Gilliland et al. (1999) report on the discovery of two high redshift SNe in the HDF (see also Mannucci \& Ferrara 1999). One of the SNe has a probable host galaxy at $z\ \sim$ 1.3 (photometrically  determined) and is likely to be a Type Ia, whereas the other SN has a probable host galaxy at $z\ \sim$ 0.95 (spectroscopically determined) and is possibly a Type II. Gilliland et al. also make detailed estimates of the expected number of Type Ia and Type II SNe in a HDF like search. With limiting magnitude I$_{AB} \sim$ 27.7 they find that $\sim$ 0.32 Type Ia SNe should be detected in a search consisting of the HDF together with an observation of the same area made two years after the HDF. Using the same cosmological model as Gilliland et al. ($\Omega_M = 0.28,\ \Omega_{\Lambda}$ = 0.72, Ho = 63.3 km/s/Mpc$^3$) and $\tau$ = 1 Gyr, we estimate $\sim$ 0.6 Type Ia SNe for a similar search. For a flat $\Omega_M$ = 1 cosmology we estimate 0.7 SNe. The main reason for the difference in results is that Gilliland et al. use a constant Type Ia SNR over the redshift range of interest (0 $< z\ \lsim$ 1.5), which is considerably lower than the mean value of our rates out to $z \sim$ 1.5.

For Type II SNe Gilliland et al. estimate $\sim$ 1.2 SNe in HDF style search. This is in good agreement with our results, even though the modeling differs in many aspects. We estimate $\sim$ 1.0 core collapse SNe for the cosmology used by Gilliland et al., and 1.3 SNe for a $\Omega_M$ = 1 cosmology.

Considering the small statistics, both estimates are consistent with the discovery of two SNe in the HDF.

Sadat et al. (1998) discuss the cosmic star formation rate, using a
spectrophotometric model for different assumptions of the dust
extinction. From this they calculate SNIa and core collapse rates, but
do not translate these into directly observable rates. Their SFR is a
factor $\gsim$ 3 higher than ours, which seems mostly to be due to the
use of different factors when converting the observed luminosity
densities to the SFRs. This also leads to higher SNRs (their Fig.
2). Sadat et al. also presents a case for Type Ia rates with a
different normalization. These rates (their Fig. 3) agree better with
our estimates at low redshifts. At high redshifts the Type Ia rates
differ more due to different modeling of these SNe.

J\o rgensen et al. (1997) attempt a calculation of the absolute rates
of Type Ia, II and Ib SNe from a population model. Although in
principle appealing, this model depends on the uncertain scenarios for
the progenitors of especially the Type Ia's, as we have already
discussed in this paper. Any estimates will therefore be sensitive to
these assumptions. They also neglect the distinction between Type
IIP's and Type IIL's, which most likely originate from different
progenitors. Further, J\o rgensen et al. assume in the calculation of
the observed magnitudes in the different bands as function of
redshift, that the spectrum is characterized by that at the peak. As
we have discussed, the spectrum and luminosity vary strongly with
time. The most serious deficiency is in our view their neglect of the
magnitude variation, as given by the light curve, which as we have
seen, changes the observed rates by large factors. Their estimates of
the observed rates are therefore highly questionable.

\section{Conclusions}
Observations of high redshift SNe are of interest for several
reasons. First of all, one has through these a direct probe of the
nucleosynthesis and star formation of the universe. In practice, there
are several obstacles for a quantitative study of these issues. The
fact that a large fraction of the star formation, and thus the SNe,
may be hidden within optically thick dust can make it difficult to
determine the total SFR and SNR accurately. This is certainly true for
the optical bands, where we have found that the predicted total number
of core collapse SNe with $z\ \lsim$ 1.5 is rather insensitive to the
assumed star formation scenario, as long as the star formation is
calculated to match the same observed luminosity densities, and the
same extinction is assumed for the UV light from the galaxies and the
SNe. Observations in the near-IR are less affected by this, and offers
a clear advantage to the observations of the far-UV, as used for Lyman
break objects. However, if a large fraction of the star formation
occurs in highly obscured star burst galaxies, also the near-IR rates
are severely affected. A further advantage of using SNe as star
formation indicators is that they are insensitive to surface
brightness selection effects. A complication when it comes to studying
the nucleosynthesis is that the yields of the supernovae may vary with
metallicity.

An important motivation for searches of SNe at high redshift is that
one can from this type of observations learn something about the SNe
themselves when observed in a different environment. In particular,
differences in the fractions of the various core collapse subclasses,
their spectra and luminosities may give new insight into the physics
of the SNe and their progenitors.

The number of core collapse SNe that can be detected with NGST, with
its expected limiting magnitude K$^\prime$ = 31.4, should be $\sim$ 45
per field in a 10$^4$ s exposure, assuming a hierarchical star
formation scenario. The mean redshift of these SNe is $<z>$ =
1.9. About one third of the SNe have $z\ \gsim$ 2.  The high dust
model results in total counts in the K$^\prime$ band that are a factor
$\sim$ 2 higher than in the hierarchical model. The estimated number
of SNe with $z\ \gsim$ 2 in the K$^\prime$ band for NGST is a factor
$\sim$ 3 higher than in the hierarchical model. The model with flat
SFR at high $z$, but with low extinction, result in a factor $\sim$ 2
higher number of SNe with $z\ \gsim$ 2, compared to the hierarchical
model, for the NGST limit.

An important practical point is that in order to
detect especially the Type IIP SNe at high $z$ it is necessary to have
a spacing between observations of $\sim$ 100 days for ground based
telescopes, and $\sim$ 1 year for deep observations with NGST. Shorter
time intervals do not allow for the luminosity of the SNe to decrease
by an amount necessary for detection.

When it comes to the observed rates of Type Ia SNe, we find that these
are highly sensitive to the star formation modeling. This is due to
the fact that the Type Ia's are less linked to the environment where
their progenitors were formed. The uncertainty in the life-time of the
progenitors, combined with the sensitivity of the Type Ia rates to the
onset of star formation in the models with a flat SFR at high $z$,
contributes to the difficulty with using Type Ia counts as probes for
either different star formation scenarios or progenitor models. This
is further hampered by the fact that even the local rate is highly
uncertain, and that this propagates to other redshifts through the
normalization of the rates at $z$ = 0. Therefore, accurate
measurements of the Type Ia rates at low $z$ are most desired.

Precise measurements of the Type Ia rates at $z\ \gsim$ 1 could
constrain the parameters to some extent. For example, in a given
cosmology there is a high redshift cutoff in the Ia rates at an epoch
that depends strongly on $\tau$, but is less dependent on the star
formation scenario.  In agreement with previous studies we find that
counts of Type Ia SNe can be used as cosmological probes. This does,
however, require that both the SFR and the unknown life time of the
Type Ia progenitors can be determined independently.

We predict the number of simultaneously detectable Type Ia SNe per
NGST field to be $\sim$ $5-25$, depending on progenitor model and star
formation scenario. Additional uncertainties widen this range even
more. Of the simultaneously observable Type Ia SNe, about 5\% are on
the rising part of the light curve. For a ground based telescopes with
limiting magnitude I$_{AB}\sim$ 27 we predict $75-400$ Type Ia's per
square degree of which $\sim$ 30\% are on the rise of the light curve.

A major technical aspect of our work is that we have tried to
incorporate as much knowledge as possible about the theoretical and
observational properties of the different classes of SNe. In
particular, we have found that the spectral evolution is important for
the magnitudes in the different bands. A striking example is the
sensitivity to the UV cutoff, which for most SNe dominate the
evolution in the optical bands. We have also found that a proper
treatment of the light curve can change the predicted rates by factors
of three or larger. An important source of uncertainty in these
estimates are the local frequencies of SNe of different classes, as
well as their distribution in luminosity. More extensive surveys with
well defined selection criteria are therefore of highest priority for
reliable predictions at high redshifts.
\begin{acknowledgements}
We are grateful to Claes-Ingvar Bj\"{o}rnsson, Ariel Goobar, Bob
Kirshner, Bruno Leibundgut, Ken Nomoto and Brian Schmidt for discussions, 
and especially
to Ron Eastman for supplying their Type IIP results in digital
form. This work is supported by the Swedish Natural Science Council
and the Swedish Board for Space Sciences.
\end{acknowledgements}
{}
\end{document}